\relax 
\documentclass[pra,preprint,amsfonts,showpacs,eqsecnum]{revtex4}         

\usepackage{latexsym}   
\usepackage[]{graphicx} 
\usepackage{times}
\usepackage{bm}

\def\defeq{\mathrel{\stackrel{{\rm def}}{=}}}   
\def\usq{\raise1pt\hbox{\texttt{\char"0D}}}  
\def\lbrk{\hfil\break\vskip-22.55pt}
\def\lbrkk{\hfil\break\vskip-45pt}

\begin{document}   

\title{Framework for quantum modeling of fiber-optical networks: Part~II
\\  (\small Rev. 0.2.5:\ \
  suggestions and corrections welcome)\vadjust{\kern20pt}}

\author{John M. Myers}

\affiliation{Gordon McKay Laboratory, Division of Engineering and Applied
Sciences\\ Harvard University, Cambridge, Massachusetts 02138}

\date{\textbf{31 May 2005}\\[30pt]}

\begin{abstract}

We formulate quantum optics to include frequency dependence in the
modeling of optical networks.  Entangled light pulses available for
quantum cryptography are entangled not only in polarization but also,
whether one wants it or not, in frequency.  We model effects of the
frequency spectrum of faint polarization-entangled light pulses on
detection statistics.  For instance, we show how polarization
entanglement combines with frequency entanglement in the variation of
detection statistics with pulse energy.

Attention is paid not only to single-photon light states but also to
multi-photon states.  These are needed (1) to analyze the dependence
of statistics on energy and (2) to help in calibrating fiber couplers,
lasers and other devices, even when their desired use is for the
generation of single-photon light.

\end{abstract} 

\maketitle
\thispagestyle{empty}

\newpage
\setcounter{page}{1}
\def\thepage{\roman{page}} 

\def\TOCsec#1#2#3{\vspace{0pt}\noindent \hangindent=.5in \noindent \hbox
to .5in{\hfil #1.\ }#2\dotfill #3\vspace{-3.5pt}}
\def\TOCsubsec#1#2#3{\noindent \hangindent=.72in \noindent \hbox to
.5in{}\hbox to .22in{#1.\ \hfil}#2\dotfill #3}
\def\TOCapp#1#2#3{\noindent \hangindent=1.04in \noindent \hbox to
.125in{}Appendix\  #1.\ $\,$#2\dotfill #3}
\def\TOCref#1#2{\noindent \hangindent=.875in \noindent \hbox to
.125in{}#1\dotfill #2}

\centerline{\large\textbf{Contents}}

\vspace{3pt} 

\noindent PART I

\TOCsec{1}{Introduction}{1}

\TOCsubsec{A}{Quantum modeling}{2} 

\TOCsubsec{B}{Aims in developing a framework}{3}

\TOCsubsec{C}{Approach}{4}

\TOCsec{2}{Modes, commutation rules, and light states}{5}

\TOCsubsec{A}{Single-photon state spread over multiple modes}{9}

\TOCsubsec{B}{Single-mode, multi-photon states}{9}

\TOCsubsec{C}{Broad-band coherent states}{10}

\TOCsubsec{D}{General state}{11}

\TOCsubsec{E}{Density matrices and traces}{12}

\TOCsubsec{F}{Partial traces of light states}{13}

\TOCsubsec{G}{Bi-photons: excitation in each of two orthogonal
modes}{15}

\TOCsec{3}{Projections}{16}

\TOCsubsec{A}{Action of single-mode projections on multi-mode
states}{17}

\TOCsubsec{B}{Multi-mode $n$-photon projector}{17}

\TOCsubsec{C}{Number operator}{18}

\TOCsec{4}{Loss and frequency dispersion}{19}

\TOCsubsec{A}{Loss cannot evade ``no cloning''}{19}

\TOCsec{5}{Local quantum fields}{20}

\TOCsubsec{A}{Temporally local hermitian fields}{21}

\TOCsubsec{B}{Time, space, and dispersion}{22}

\TOCsubsec{C}{Projections in terms of local operators}{23}

\TOCsec{6}{Scattering matrix}{23}

\TOCsubsec{A}{Network without frequency mixing}{24}

\TOCsec{7}{Polarized and entangled light states}{25}

\TOCsubsec{A}{Fiber splice (without extraneous modes)}{26}

\TOCsubsec{B}{Coupler}{27}

\TOCsubsec{C}{Entangled states}{27} 

\TOCsubsec{D}{Polarization-entangled states}{28}

\TOCsec{8}{Detection}{29}
  
\TOCsubsec{A}{Simple examples}{30}
 
\TOCsubsec{B}{Model of APD detector for quantum cryptography}{31}

\TOCsubsec{C}{Detection probabilities}{33}

\TOCsubsec{D}{Effect of time bounds on detection}{36}

\TOCsubsec{E}{Detection, energy, and photon subspaces}{36}

\TOCsubsec{F}{Preceding the APD detector by a beam-splitter}{37}

\TOCsec{9}{Polarization-entangled light for QKD}{39}

\TOCsubsec{A}{Bi-photon light states}{39}

\TOCsubsec{B}{Effect of a beam splitter}{41}

\TOCsubsec{C}{Effect of polarization rotation}{41}

\noindent PART II

\TOCsec{10}{Modeling polarization-entangled QKD}{43} 

\TOCsubsec{A}{Outcomes and probabilities}{45}

\TOCsubsec{B}{Light state}{48}

\TOCsubsec{C}{Energy profile}{50}

\TOCsubsec{D}{Calculation of probabilities}{51}

\TOCsubsec{E}{Case I: No frequency entanglement}{56} 

\TOCsubsec{F}{Case II: Limit of extreme frequency entanglement as
$\zeta\to \pm \infty$}{58}

\TOCsubsec{G}{Example numbers}{59}

\TOCapp{A}{Background}{59}

\TOCapp{B}{Operator Lemmas}{60}

\TOCapp{C}{Algebra of frequency-entangled operators}{64}

\TOCapp{D}{Fourier transforms in space and time}{76}

\TOCapp{E}{Expansion of light states in tensor products of 
broad-band\vadjust{\kern-7pt}\hfil\break coherent states}{77}

\TOCapp{F}{MATLAB programs for Section \protect\ref{sec:10}}{78}

\TOCref{References}{97}    

\newpage\clearpage

\def\thepage{\arabic{page}}  

\setcounter{page}{43}
\centerline{\large\bf PART\, II} 
\setcounter{section}{9}
\setcounter{figure}{4}

\section{Modeling polarization-entangled QKD}\label{sec:10}

\begin{figure}[t]
\begin{center}
\includegraphics[width=5.5in]{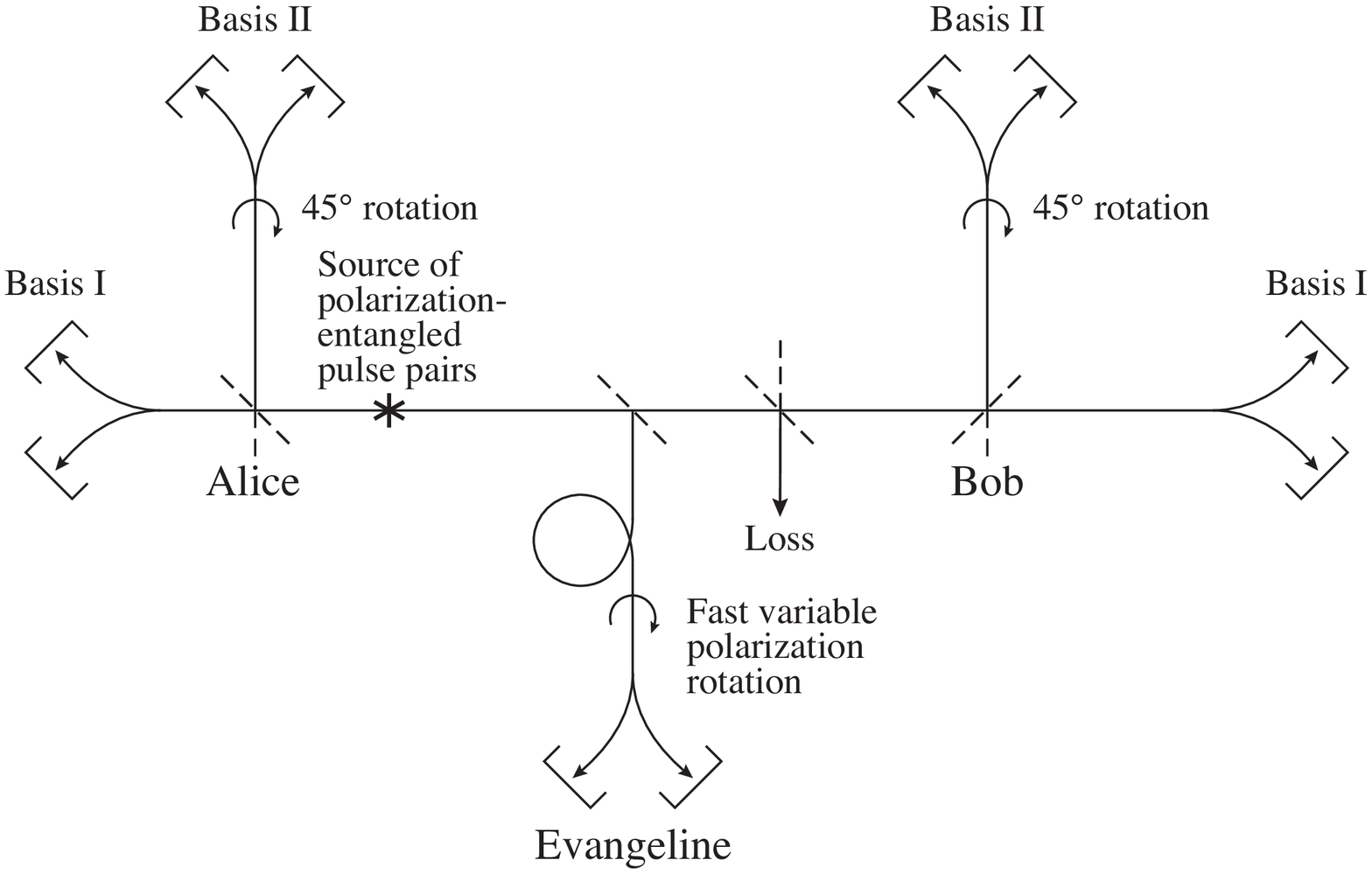}
\end{center}
\vspace{-20pt}
\caption{Polarization-entangled QKD system subjected to
eavesdropping attack.}\label{fig:5}
\end{figure}

The language offered in the preceding sections supports a wide variety
of models of the system shown in Fig.\ \ref{fig:5} \cite{entangQKD}, as
well as of many other systems.  Here we offer a first-cut model of a
fiber-optic network that employs polarization-entangled light for quantum
key distribution (QKD).  At the present stage of development of quantum
key distribution, the purpose of the modeling can hardly be to replace
experiments: we lack convincing reasons, whether theoretical or
experimental, on which to ground the guesswork necessary to generate
numbers.  Rather, we show how a model drawing on some questionable
guesses can stimulate experiments.

Specifically, the model offered shows that if one assumes a Poisson
distribution and one assumes a light state invariant under identical
SU(2) transformations of the light to both Alice and Bob, then such
and such relations hold between energy, Bob's error rate, and
Evangeline's entropy.  These relations are the ``conclusion part'' of a
statement that includes also an ``if part.''  As remarked by Dave
Pearson, any interesting conclusion makes the ``if part'' worth
exploring, for instance by means of experiments.  What evidence can we
find in the lab for or against the assumption of SU(2) invariance?
For or against a Poisson distribution of photon number?  Expecting to
later challenge some of our own assumptions, we try to make our
modeling modular, so that the assumptions can be changed, both to make
room in the future for improvement, and to lay the ground for studies
of sensitivity to these assumptions.

Now to business. Consider a simplified experiment to explore the
relation between the energy of a polarization-entangled light pulse and
various detection probabilities relevant to: (1) quantum bit error rate
(QBER), (2) sifted bit rate, and (3) an eavesdropper's entropy.  We start
by being interested in the QKD topology shown in Fig.\ \ref{fig:5}, in
which Alice and Bob each operate passively to detect in two bases, with
two detectors per basis.

For this first model we assume: 
\begin{enumerate}
\item An ensemble of trials, one entangled pulse pair per trial.
\item Pulses timed and detectors managed so that trapping, dead time,
and other memory effects of detection are negligible.
\item \textbf{Light state} leaving transmitter, propagating in modes
$a_1,\,a_2$ to Alice and in modes $\tilde{b}_1,\,\tilde{b}_2$ toward Bob,
invariant under application of any given SU(2) transformation to both
the $a$-modes and the $b$-modes, as discussed below.
\item A probability of 1/2 for Alice and Bob having matching bases.
\item Alice and Bob use ``on-off'' detectors as described in
Sec.\ \ref{sec:8}; the detector for mode $a_j$ has a dark-count
probability
$p_{\rm dark}(a_j)$ and an efficiency $\eta_{\rm det}(a_j)$.
\item A fraction $\eta_{\rm trans}$ of the energy transmitted
to Bob survives attenuation, as described by frequency-independent
coupling to an undetected (undesired) mode.
\item Poisson distribution of photon number in the energy transmitted
to Bob.
\item \textbf{Sifting rule}: Alice and Bob discard a bit except when (a) 
they each get one and only one detection, and (b) their
bases match.
\item\textbf{Eavesdropping attack}: Evangeline sneaks a
non-polarizing beam splitter into Bob's fiber to siphon off her
choice of a fraction of the light into modes $b_3$ and $b_4$ that
propagate in her fiber.  This runs through a long, lossless, delay
line to a rapidly variable polarization rotator, followed by a pair
of perfect photon-number detectors.  The delay line allows Evangeline
to postpone detection until she has learned what basis Bob has used
for his detection; she then rotates or not, as necessary to choose
the basis that matches Bob's.
\end{enumerate}

\noindent Much of the analysis is independent of assumptions 5 and 6; we
indicate later where these assumptions enter.

\begin{figure}[t]
\begin{center}
\includegraphics[width=5.3in]{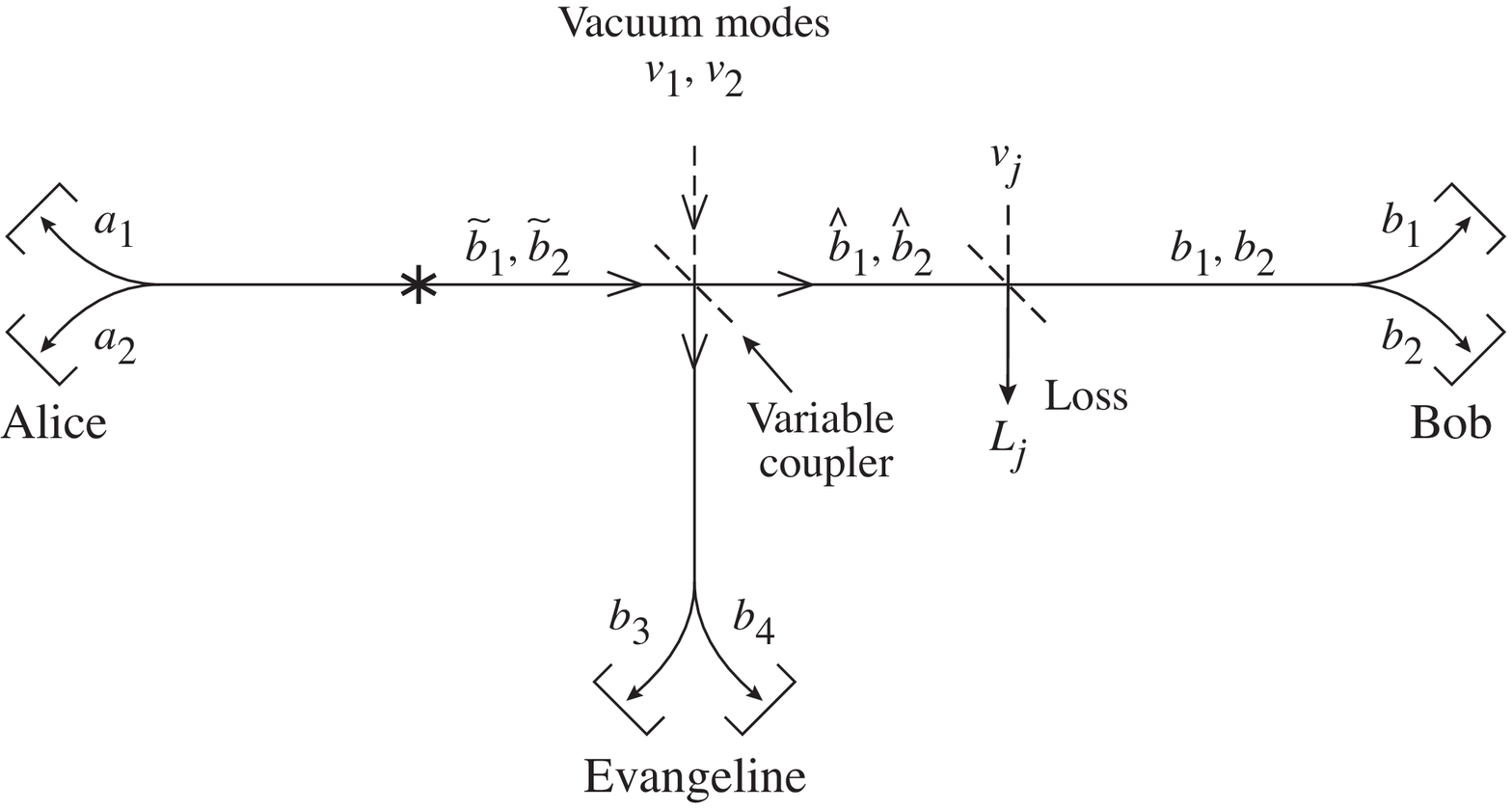}
\end{center}
\vspace{-24pt}
\caption{Simplified network.}\label{fig:6}
\end{figure}

To analyze the assumed eavesdropping attack, in which our attacker
Evangeline knows Bob's basis, we do not need the whole setup of
Fig.\ \ref{fig:5}; we can simplify by leaving out the rotated
bases, and the two beam splitters that support them, along with
Evangeline's variable polarization rotator.  This results in
Fig.\ \ref{fig:6}.  One can think of the modes with subscripts 1
and 3 as `vertically polarized' and those with subscripts 2 and 4 as
`horizontally polarized.'  We model the variable coupler by which
Evangeline taps off energy by an SU(2) transformation.  For $j = 1,2$
we have
\begin{equation}
\left[\begin{array}{l}b_j(\omega) \\ b_{j+2}(\omega)\end{array}\right]
= \left[\begin{array}{cc}u & -v^* \\v & u^*\end{array}\right]
\left[\begin{array}{l}\tilde{b}_j(\omega) \\
v_j(\omega)\end{array}\right],
\label{eq:coupler}
\end{equation}
where the $v_j$ are vacuum modes assumed unexcited, and $|u|^2 + |v|^2
=1$.  Inverting this equation, we find 
\begin{equation}
\tilde{b}_j(\omega) = u^*b_j(\omega) + v^*b_{j+2}(\omega).
\label{eq:detMode}
\end{equation}

\subsection{Outcomes and probabilities}
By an {\em elementary outcome} we mean a possible joint response of
all the detectors involved.  We view an elementary outcome as
constituted from {\em components}, one component for each detector.
In the context of the model presented here, an elementary outcome
consists of a bit string for the modes for which detection is binary
(such as APD detectors as modeled in Sec.\ \ref{sec:8}), and a
non-negative integer for each of Evangeline's two modes (that we imagine
as having photon-counting detectors).  Each possible joint response of the
binary detectors can be expressed by partitioning the set of modes
subject to binary detection into a set $\mathbf{J}_0$ for which the
response is `no-detect' and a set $\mathbf{J}_1$ for which the
response is `detect'; correspondingly any elementary outcome has the
form $(\mathbf{J}_0,\mathbf{J}_1,k,m)$ where $k = N(b_3)$ and $m =
N(b_4)$.  The corresponding detection operator factors; given any
normalized state vector $|\psi\rangle$,
\begin{equation}
\Pr(\mathbf{J}_0,\mathbf{J}_1,k,m) = \langle
\psi|\mathbf{M}(\mathbf{J}_0,\mathbf{J}_1,k,m)|\psi \rangle,
\label{eq:firstProb}
\end{equation}
with 
\begin{equation}
\mathbf{M}(\mathbf{J}_0,\mathbf{J}_1,k,m) =
P_k(b_3)P_m(b_4)\left(\prod_{x\in \mathbf{J}_0}M_0(x)\right)\left(
\prod_{x \in \mathbf{J}_1}M_1(x)\right),
\label{eq:M1prodkm}
\end{equation}
where $P_k(b_3)$ and $P_m(b_4)$ are projections and $x$ ranges
over modes in the lists $\mathbf{J}_0$ and $\mathbf{J}_1$.

When we want to ignore Evangeline's detections, we have a
non-elementary outcome $(\mathbf{J}_0,\mathbf{J}_1)$ with the
corresponding probability
\begin{equation}
\Pr(\mathbf{J}_0,\mathbf{J}_1) = \sum_{k,m=0}^\infty
\Pr(\mathbf{J}_0,\mathbf{J}_1,k,m).
\end{equation}
In expressions for outcomes that enter calculations we will often
write, in place of $(\mathbf{J}_0,\mathbf{J}_1)$, a list of all the
modes in these sets in the order $(a_1\,a_2\,b_1\,b_2)$ with a bar
placed over the (undetected) modes that belong to $\mathbf{J}_0$.
Thus an outcome specified by $\mathbf{J}_0 = \{a_2,b_1\}$ and
$\mathbf{J}_1= \{a_1,b_2\}$ will also be written
$(a_1\,\overline{a}_2\,b_1\,\overline{b}_2)$. 

Only four such outcomes survive the assumed sifting rule that requires
exactly one detection by Alice and exactly one detection by Bob.
Because of the assumed light state, Bob takes `detect' on $b_1$ to be
a 1-bit for a quantum key, while Alice takes `detect' not on $a_1$ but
on $a_2$ for a 1-bit, etc.  With this rule the four outcomes that
survive sifting are related to `correct bit' and `error' as follows:
$(a_1\,\overline{a}_2\,\overline{b}_1\,b_2)$ and 
$(\overline{a}_1\,a_2\,b_1\,\overline{b}_2)$ are correct from the
standpoint of QKD, while $(\overline{a}_1\,a_2\,\overline{b}_1\,b_2)$ and
$(a_1\,\overline{a}_2\,b_1\,\overline{b}_2)$ are errors.

The probability of a bit surviving sifting for cases in which the bases
match is, in this simplified model,
\begin{eqnarray}\lefteqn{
\Pr(\mbox{trial produces a sifted bit})}\quad \nonumber \\ & =&
\Pr(\overline{a}_1\,a_2\,\overline{b}_1\,b_2) +
\Pr(a_1\,\overline{a}_2\,b_1\,\overline{b}_2) +
\Pr(\overline{a}_1\,a_2\,b_1\,\overline{b}_2) +
\Pr(a_1\,\overline{a}_2\,\overline{b}_1\,b_2).
\end{eqnarray}
The probability of an error in a sifted bit is
\begin{equation}
\Pr(\mbox{bit error$|$sifted}) =
\frac{\Pr(\overline{a}_1\,a_2\,\overline{b}_1\,b_2) +
\Pr(a_1\,\overline{a}_2\,b_1\,\overline{b}_2)}{\Pr(\mbox{trial
produces a sifted bit})}.
\end{equation}
For trials that result in error-free bits, we want to know the degree
to which Evangeline's outcomes $N(b_3)$ and $N(b_4)$ leave her
ignorant concerning Bob's outcomes.  This ignorance of Evangeline with
respect to error-free bits is measured by R\'enyi entropy of order
$R$. This $R$-entropy depends on the conditional probability
that Bob received a 1, given Evangeline's detector response $(k,m)$.
This we denote
\begin{eqnarray} 
\mbox{Ev}(k,m)& \defeq  &
\Pr(\overline{a}_1\,a_2\,b_1\,\overline{b}_2;k,m\,|\{\overline{a}_1\,a_2\,
b_1\,\overline{b}_2;k,m\} \mbox{ or
}\{a_1\,\overline{a}_2\,\overline{b}_1\,b_2;k,m\}) \nonumber \\[12pt] 
& = & \frac{\Pr(\overline{a}_1\,a_2\,b_1\,\overline{b}_2;k,m)}
{\Pr(\overline{a}_1\,a_2\,b_1\,\overline{b}_2;k,m)+
\Pr(a_1\,\overline{a}_2\,\overline{b}_1\,b_2;k,m)}.
\label{eq:EvProb}
\end{eqnarray}
Evangeline's $R$-entropy given  $N(b_3) = n$ and $N(b_4) = m$
is 
\begin{equation}
\mbox{Ent}_R(k,m) = \frac{1}{1-R}[(\mbox{Ev}(k,m))^R +
(1-\mbox{Ev}(k,m))^R].
\end{equation}
Evangeline's average $R$-entropy on error-free bits is then
\begin{equation}
\mbox{AvEnt}_R = \frac{\sum_{k,m = 0}^\infty
[\Pr(\overline{a}_1\,a_2\,b_1\,\overline{b}_2;k,m)+
\Pr(a_1\,\overline{a}_2\,\overline{b}_1\,b_2;k,m)]\mbox{Ent}_R(k,m)}{
\Pr(\overline{a}_1\,a_2\,b_1\,\overline{b}_2)+
\Pr(a_1\,\overline{a}_2\,\overline{b}_1\,b_2)}.\label{eq:EvAvRenEFB}
\end{equation}

Altogether there are six types of outcomes: those with and without the
distinction ``$k,m$'' for Evangeline's detectors, and for each of these
the general case of an arbitrary energy distribution and two special
cases of the Poisson energy distribution and that of a single photon
number $n$.  Of course all these probabilities (and the $R$-entropy)
depend on both the light state $|\psi\rangle$ and the parameters of the
APD model of Alice's and Bob's detectors.

\subsection{Light state}
We formulate a family of states for entangled light, as is discussed in
more detail in Appendix~\ref{app:C}.  The calculations are complicated;
here we carry them out for two limiting cases that are relatively simpler.

To begin rather generally, we are concerned with an otherwise
arbitrary normalized state vector
\begin{equation}
|\psi\rangle = \sum_{n=0}^\infty C_n |\psi_n \rangle,
\label{eq:stateStart}
\end{equation}
where the state $|\psi_n\rangle$ signifies $n$ photons transmitted
to Bob and
\begin{equation}
\sum_{n=0}^\infty |C_n|^2 = 1.  
\end{equation}
The index $n$ for `photon number' to Bob (which here will be the same
as that for Alice) takes specific meaning when we assume a
polarization-entangled light pulse invariant under matching SU(2)
transforms of both $a$-modes and $\tilde{b}$-modes, for which
\begin{equation}
 |\psi_n\rangle = \mathcal{N}(g_\zeta,n)
[g_\zeta\!:\!(a_1\tilde{b}_2-a_2\tilde{b}_1)^\dag]^n|0\rangle ,
\label{eq:state}
\end{equation}
with
\begin{equation}
g_\zeta\!:\!(a_1\tilde{b}_2-a_2\tilde{b}_1)^\dag
\defeq  \int_{-\infty}^\infty d\omega_1\,d\omega_2\,
g_\zeta(\omega_1,\omega_2)[a_1^\dag(\omega_1)\tilde{b}_2^\dag(\omega_2)
  - a_2^\dag(\omega_1)\tilde{b}_1^\dag(\omega_2)],
\end{equation}
where we assume a family of functions $g_\zeta(\omega,\tilde{\omega})$ of
the following form.  For any real-valued functions $\phi(\omega)$
and $\tilde{\phi}(\tilde{\omega})$ and positive real parameters{
$\sigma$ and $\tilde{\sigma}$, let
\begin{equation}
g_\zeta(\omega,\tilde{\omega}) =
\frac{1}{\sqrt{\sigma\tilde{\sigma}}}\,e^{i\phi(\omega)}
e^{i\tilde{\phi}(\tilde{\omega})} F\left(\zeta;\frac{\omega-
\omega_0}{\sigma},\frac{\tilde{\omega}-\tilde{\omega}_0}
{\tilde{\sigma}}\right),
\label{eq:gzetapp}
\end{equation}
where we define
\begin{eqnarray}
F(\zeta;x,y) & \defeq  &
\sqrt{\frac{2}{\pi}}\exp\left\{-\frac{1}{2}\left[\left(\sqrt{\zeta^2+1}
+\zeta\right) (x+y)^2 +
\left(\sqrt{\zeta^2+1}+\zeta\right)^{-1}(x-y)^2\right]\right\}
\nonumber
\\ & = &
\sqrt{\frac{2}{\pi}}\exp\left\{-\left[\sqrt{\zeta^2+1}\,(x^2 + y^2) + 2
\zeta x y\right]\right\};
\label{eq:FDef}
\end{eqnarray}
regardless of the value of $\zeta$,
\begin{equation}
\int^\infty_{-\infty} dx\,dy\,|F(\zeta;x,y)|^2 = 1.
\end{equation}
Thus for any choice of center frequencies $\omega_0$ and
$\tilde{\omega}_0$, bandwidth parameters $\sigma$ and
$\tilde{\sigma}$, and phase functions $\phi(\omega)$ and
$\tilde{\phi}(\tilde{\omega})$, we get a family of $g_\zeta$'s.

In Eq.\ (\ref{eq:state}), $\,\mathcal{N}(g_\zeta,n)$ is a normalization
constant that makes $|\psi_n\rangle$ have unit norm, so that it is
defined by
\begin{equation}
{}[\mathcal{N}(g_{\zeta},n)]^2 = \langle
0|[g_\zeta^*\!:\!(a_1\tilde{b}_2-a_2\tilde{b}_1)]^n
[g_\zeta\!:\!(a_1\tilde{b}_2-a_2\tilde{b}_1)^\dag]^n|0\rangle^{-1}.
\label{eq:calN}
\end{equation}
Writing $(a_jb_k)$ for
$g^*_\zeta\!:\!a_jb_k$, we have from
Eq.\ (\ref{eq:calN})
\begin{eqnarray}
{}[\mathcal{N}(g_\zeta,n)]^2 & = & \langle
0|[(a_1\tilde{b}_2)-(a_2\tilde{b}_1)]^n
[(a_1\tilde{b}_2)-(a_2\tilde{b}_1)]^{\dag n}|0\rangle^{-1} \nonumber
\\ & = & \langle 0|\sum_{k=0}^n\left(\begin{array}{c}n\\[-5pt] k
\end{array}
\right)(a_1\tilde{b}_2)^k (-a_2\tilde{b}_1)^{n-k}\sum_{\ell=0}^n
\left(\begin{array}{c}n\\[-5pt] \ell\end{array} \right)
(a_1\tilde{b}_2)^{\dag \ell}(-a_2\tilde{b}_1)^{\dag
(n-\ell)}|0\rangle^{-1} \nonumber \\ & = &
\left\{\sum_{k=0}^n\left(\begin{array}{c}n\\[-5pt] k\end{array}
\right)^2\langle 0|(a_1\tilde{b}_2)^k (a_2\tilde{b}_1)^{n-k}
(a_1\tilde{b}_2)^{\dag k}(a_2\tilde{b}_1)^{\dag
(n-k)}|0\rangle\right\}^{-1}\nonumber \\
& = & \left\{n!\sum_{k=0}^n\left(\begin{array}{c}n\\[-5pt] k\end{array}
\right)\Xi_{g_\zeta}(k)\Xi_{g_\zeta}(n-k)\right\}^{-1},
\label{eq:cal2N}
\end{eqnarray}
where we define
\begin{equation}
\Xi_g(n) \defeq  \frac{1}{n!}\,\langle
0|(g_\zeta^*\!:\!ab)^n(g_\zeta\!:\!a^\dag b^\dag)^n|0\rangle,
\label{eq:XiDef}
\end{equation}
and the last equality in Eq.\ (\ref{eq:cal2N}) comes from
tensor-product factoring. Note that all that matters
about $a(\omega)$ and $b(\omega)$ in this definition is that
they are mutually orthogonal; any other pair would give the
same value.

Re-expressing
$|\psi_n\rangle$ in terms of detector modes per Eq.\
(\ref{eq:detMode}), one has for the light state
\begin{equation}
|\psi_n\rangle  =  
\mathcal{N}(g_\zeta,n)[g_\zeta\!:\!(ua_1^\dag b_2^\dag + va_1^\dag
  b_4^\dag - u a_2^\dag b_1^\dag - va_2^\dag b_3^\dag)]^n|0\rangle.
\label{eq:psi2}
\end{equation}
When we write $(a_j^\dag b_k^\dag)$ as shorthand for
$g_\zeta\!:\!a_j^\dag b_k^\dag$, this becomes
\begin{equation}
|\psi_n\rangle = \sum_{{\scriptstyle j,k,\ell,m = 0}\atop
{\scriptstyle j+k+\ell+m = n}}^n \mathcal{N}(g_\zeta,n)\,
  \frac{n!}{j!k!\ell!m!}\,u^{j+\ell}v^{k+m}(-1)^{\ell+m}(a_1^\dag
  b_2^\dag)^j(a_1^\dag b_4^\dag)^k(a_2^\dag b_1^\dag)^\ell(a_2^\dag
  b_3^\dag)^m|0\rangle.
\label{eq:psiNDef}
\end{equation} 

\subsection{Energy profile}
We suppose that the light is generated by equipment close to Alice, so
that the energy exposed to eavesdropping is in the $\tilde{b}$-modes
rather than in the $a$-modes.  For a state of the form defined by
Eqs.\ (\ref{eq:stateStart}), (\ref{eq:state}), we want to express the
expectation energy for modes $\tilde{b}_1$ and $\tilde{b}_2$, denoted
\begin{equation}
\mbox{Energy}(\tilde{b}_1,\tilde{b}_2) =
\langle \psi|H(\tilde{b}_1,\tilde{b}_2)|\psi\rangle,
\end{equation}
where for present purposes we approximate the hamiltonian operator
\begin{equation}
H(\tilde{b}_1,\tilde{b}_2) = \hbar \int_{-\infty}^\infty
  d\omega\,|\omega|[\tilde{b}_1^\dag(\omega)\tilde{b}_1(\omega)+
  \tilde{b}_2^\dag(\omega)\tilde{b}_2(\omega)]
\end{equation} for narrow-band signals by
\begin{equation}
H(\tilde{b}_1,\tilde{b}_2) \approx \hbar \omega_0\int_{-\infty}^\infty
  d\omega\,[\tilde{b}_1^\dag(\omega)\tilde{b}_1(\omega)+
  \tilde{b}_2^\dag(\omega)\tilde{b}_2(\omega)],
\end{equation} where $\omega_0$ is the carrier angular frequency,
similar to that of Eq.\ (\ref{eq:centerF}).  The commutation relation
of Lemma (\ref{eq:commsing}) of Appendix~\ref{app:B} leads then to
\begin{equation}
\mbox{Energy}(\tilde{b}_1,\tilde{b}_2) =
\hbar \omega_0\sum_{n=1}^\infty n|C_n|^2.  
\label{eq:energyC}
\end{equation}
Denote the `mean photon number' by 
\begin{equation}
\mu \defeq  \frac{1}{\hbar
\omega_0}\mbox{Energy}(\tilde{b}_1,\tilde{b}_2),
\end{equation}
so that, from Eq.\ (\ref{eq:energyC}) we have
\begin{equation}
\mu = \sum_{n=1}^\infty n|C_n|^2.
\end{equation}

To produce a dependence of probabilities on $\mu$, we have to choose
an energy profile, which means choosing the $C_n$.  As a first cut, we
will show consequences of assuming a Poisson distribution
\begin{equation}
\mbox{Assume:}\quad |C_n|^2 = \frac{e^{-\mu}\mu^n}{n!}\,.
\label{eq:Cassume}
\end{equation}

\subsection{Calculation of probabilities}

Lacking strong theoretical or experimental evidence to guide the
choice of energy distribution of light for QKD, we arrange the
modeling so that this distribution can be entered as a parameter.  To
this end we provide for modeling the contribution of individual values
of photon number $n$.  As remarked earlier, we want probabilities
$\Pr(\mathbf{J}_0,\mathbf{J}_1,k,m)$ and
$\Pr(\mathbf{J}_0,\mathbf{J}_1)$, i.e., with and without the
``$k,m$'' distinction; further we want each of these for the general
case of an energy distribution $\mathbf{C} = \{|C_n|^2\}$ and for the
two special cases of (1) a Poisson distribution and (2) an $n$-photon
state.  Altogether, this makes 2 $\times$ 3 = 6 types.  Each of these
six types will be expressed by a corresponding function $\mathcal{T}$;
we will soon see $\mathcal{T}_{\mathbf{C},km}$,
$\mathcal{T}_{\mu,km}$, $\mathcal{T}_{n,km}$,
$\mathcal{T}_{\mathbf{C}}$, $\mathcal{T}_\mu$, and $\mathcal{T}_n$.
These $\mathcal{T}$ (for ``total'') functions will be calculated
as sums of corresponding functions $\mathcal{F}$ that are decorated
with the same subscripts.

To start with, for purposes of calculating probabilities we break the
state $|\psi_n\rangle$ down further in terms of the response of
Evangeline's photon-number detectors, assumed expressed by projection
operators for modes $b_3$ and $b_4$:
\begin{equation}
|\psi_n\rangle = \sum_{{\scriptstyle k,m = 0}\atop {\scriptstyle k+m \le n}}^n |\psi_{n,km}\rangle,
\end{equation}
where
\begin{eqnarray}
|\psi_{n,km}\rangle &\defeq & 
 P_k(b_3)P_m(b_4)|\psi_n\rangle\nonumber \\ & =&
 \frac{\mathcal{N}(g_\zeta,n)n!}{k!m!} \sum_{j=0}^{n-k-m}
 \frac{u^{n-k-m}v^{k+m}}{j!(n-k-m-j)!}\nonumber\\
&&{}\times(-1)^{n-k-j}(a_1^\dag
 b_2^\dag)^j(a_1^\dag b_4^\dag)^k(a_2^\dag
 b_1^\dag)^{n-k-m-j}(a_2^\dag b_3^\dag)^m|0\rangle.\quad
\label{eq:psikmDef}
\end{eqnarray}
$|\psi\rangle$ and $|\psi_n\rangle$ but not $|\psi_{n,km}\rangle$ have
unit norm
\begin{equation}
\sum_{k,m=0}\langle \psi_{n,km}|\psi_{n,km}\rangle =
\langle\psi_n|\psi_n\rangle = 1.
\end{equation}

As shorthand for products of detection operators we write
\begin{eqnarray}
\mathbf{M}_0(\mathbf{J}_0) & = & \prod_{x\in \mathbf{J}_0}M_0(x),
\nonumber \\
\mathbf{M}_1(\mathbf{J}_1) & = &
\prod_{x \in \mathbf{J}_1}M_1(x).
\label{eq:M1prod}
\end{eqnarray}

The detection operators for Alice and Bob assumed here ``respect'' the
numbers $n$, $k$, and $m$ in the sense that, for any generic $M$ of
these operators,
\begin{eqnarray}
\langle\psi|M|\psi\rangle & = & \sum_{n=0}^\infty
|C_n|^2\langle\psi_n|M|\psi_n\rangle, \nonumber \\
\langle\psi_n|M|\psi_n\rangle & = & \sum_{{\scriptstyle
k,m=0}\atop{\scriptstyle k+m \le
  n}}^n\langle\psi_{n,km}|M|\psi_{n,km}\rangle.
\end{eqnarray}
Thus from Eq.\ (\ref{eq:firstProb}) we get, putting all this together,
\begin{equation}
\Pr(\mathbf{J}_0,\mathbf{J}_1,k,m) = \sum_{n = k+m}^\infty|C_n|^2
  \langle\psi_{n,km}|\mathbf{M}_0(\mathbf{J}_0)\mathbf{M}_1(\mathbf{J}_1)
  |\psi_{n,km}\rangle,
\label{eq:probEn}
\end{equation}
where we adopt the convention that $|\psi_{n,km}\rangle = 0$ if $k+m
>n$.  For probabilities that are indifferent to Evangeline's outcome
components, we have 
\begin{equation}
\Pr(\mathbf{J}_0,\mathbf{J}_1) = \sum_{n=0}^\infty|C_n|^2
\sum_{{\scriptstyle k,m =0}\atop{\scriptstyle k+m\le n}}^n
  \langle\psi_{n,km}|\mathbf{M}_0(\mathbf{J}_0)\mathbf{M}_1(\mathbf{J}_1)
  |\psi_{n,km}\rangle.
\label{eq:probEnn}
\end{equation}

This calculation is centered on
\begin{equation}
\langle\psi_{n,km}|\mathbf{M}_0(\mathbf{J}_0)\mathbf{M}_1(\mathbf{J}_1)
|\psi_{n,km}\rangle. 
\label{eq:mainform}
\end{equation}
{}From Eq.\ (\ref{eq:formM0}) we have
\begin{equation}
\langle\psi_{n,km}|\mathbf{M}_0(\mathbf{J}_0)\mathbf{M}_1(\mathbf{J}_1)
|\psi_{n,km}\rangle = \mathcal{T}_{n,km}(\mathbf{J}_0,\mathbf{J}_1),
\label{eq:mainEval}
\end{equation}
where we define 
\begin{equation}\mathcal{T}_{n,km}(\mathbf{J}_0,\mathbf{J}_1)
\defeq (-1)^{\#(\mathbf{J}_0)}\sum_{\mathbf{X} \subset
\mathbf{J}_1} \mathcal{F}_{n,km}(\mathbf{J}_0\|\mathbf{X}),
\label{eq:calTDef}
\end{equation}
with $\mathcal{F}_{n,km}$ defined by
\begin{equation}
\mathcal{F}_{n,km}(\mathbf{J}_0\|\mathbf{X})\defeq 
(-1)^{\#(\mathbf{J}_0\|\mathbf{X})}
\langle\psi_{n,km}|\mathbf{M}_0(\mathbf{J}_0\|\mathbf{X})
|\psi_{n,km}\rangle.
\label{eq:formM0ch10}
\end{equation}
(Note that the sum is
over all subsets of $\mathbf{J}_1$, including both $\mathbf{J}_1$
itself and the empty set $\phi$.) 

In our numerical programs it proved convenient to code the arguments
$\mathcal{F}(\mathbf{J}_0\|\mathbf{X})$ (of any type of
$\mathcal{F}$-function) by a 4-bit vector ordered by all the modes
$a_1,a_2,b_1,b_2,$ with a 1 if the mode belongs to
$\mathbf{J}_0\|\mathbf{X}$ and zero otherwise.
Thus $\mathcal{F}(\overline{a}_1,\overline{a}_2,\overline{b}_1)$ would
be coded as $\mathcal{F}(1110)$.
For example, for one of the probabilities
that enter Eq.\ (\ref{eq:EvProb}), we have
\begin{eqnarray}
\lefteqn{\mathcal{T}_{n,km}(\overline{a}_1\,a_2\,b_1\,\overline{b}_2)
}\quad \nonumber \\ &  =  &
  \mathcal{F}_{n,km}(\overline{a}_1,\overline{b}_2) +
  \mathcal{F}_{n,km}(\overline{a}_1,\overline{b}_2,\overline{a}_2) +
  \mathcal{F}_{n,km}(\overline{a}_1,\overline{b}_2,\overline{b}_1)
  +\mathcal{F}_{n,km}(\overline{a}_1,\overline{b}_2,\overline{a}_2,
  \overline{b}_2)\qquad\nonumber \\
  & = &
  \mathcal{F}_{n,km}(1 0 0 1) + \mathcal{F}_{n,km}(1 1 0 1) +
  \mathcal{F}_{n,km}(1 0 1 1) +\mathcal{F}_{n,km}(1 1 1 1).
\end{eqnarray}
This is convenient because of a trick of using a second coding scheme
for coding the argument of the $\mathcal{T}$-function: The
$\mathcal{T}$-function argument is coded (`negatively' so to speak) by
assigning a 1 if the mode appears with a bar over it and a 0 otherwise.
With these two coding schemes, the code for the $\mathcal{T}$-function
argument becomes the code for the first $\mathcal{F}$-function
argument, and the rest of the $\mathcal{F}$-function arguments are
obtained by ``filling in zeros'' in all possible ways.

In this way one evaluates Eq.\ (\ref{eq:mainform}) using only $M_0$
operators.  Drawing on the prescription of Proposition
(\ref{eq:probEff}), to calculate $\mathcal{F}_{n,km}$ we define
\begin{equation}|\phi_{n,km}(\alpha_1,\alpha_2,\beta_1,\beta_2)\rangle,
\end{equation}
the vector obtained from the expression in Eq.\
(\ref{eq:psikmDef}) for $|\psi_{n,km}\rangle$ by replacing, for $j =
1,\,2$,
$a^\dag_j(\omega)$ by $\alpha_j^{1/2}a^\dag_j(\omega)$ and
$b^\dag_j(\omega)$ by $\beta_j^{1/2}b^\dag_j(\omega)$.
This substitution yields
\begin{eqnarray}\lefteqn{
\langle
\phi_{n,km}(\alpha_1,\alpha_2,\beta_1,\beta_2)|\phi_{n,km}
(\alpha_1,\alpha_2,\beta_1,\beta_2)\rangle }\quad\nonumber \\[4pt]
 &= &
 \frac{|\mathcal{N}(g_\zeta,n)|^2n!^2}{k!^2m!^2}\, \sum_{j=0}^{n-k-m}\,
 \frac{|u|^{2(n-k-m)}(1-|u|^2)^{k+m}}{j!^2(n-k-m-j)!^2}\,
\alpha_1^{j+k}\alpha_2^{n-k-j}\beta_1^{n-k-m-j}\beta_2^j\,
\mathbf{\chi},\nonumber\\
\end{eqnarray}
where we define 
\begin{eqnarray}
\mathbf{\chi} & = & \langle 0 |(a_1 b_2)^j(a_1 b_4)^k(a_2
b_1)^{n-k-m-j} (a_2 b_3)^m |(a_1^\dag b_2^\dag)^j(a_1^\dag
b_4^\dag)^k(a_2^\dag b_1^\dag)^{n-k-m-j} (a_2^\dag
b_3^\dag)^m|0\rangle \nonumber \\ & = & \langle 0|(a_1 b_2)^j(a_1
b_4)^k (a_1^\dag b_2^\dag)^j (a_1^\dag b_4^\dag)^k|0\rangle \,\langle
0|(a_2 b_1)^{n-k-m-j}(a_2 b_3)^m (a_2^\dag b_1^\dag)^{n-k-m-j}
(a_2^\dag b_3^\dag)^m|0\rangle, \nonumber \\ &&
\label{eq:chiDef}
\end{eqnarray}
with the factorization coming from a tensor product.
{}From Proposition} (\ref{eq:nkrelation}) we have
\begin{equation}
\langle 0|(a_1 b_2)^j(a_1 b_4)^k (a_1^\dag b_2^\dag)^j (a_1^\dag
b_4^\dag)^k|0\rangle = \frac{j!k!}{(j+k)!}\langle 0|(a b)^{j+k} (a^\dag
b^\dag)^{j+k}|0\rangle = j!k!\,\Xi_{g_\zeta}(j+k),
\label{eq:targ2}
\end{equation}
where, as in Appendix~\ref{app:C}, we define 
\begin{equation}
\Xi_{g_\zeta}(n) \defeq  \frac{1}{n!}\,\langle 0|(a
b)^n (a^\dag b^\dag)^n|0\rangle.
\end{equation}
As a result, we have
\begin{equation}
\mathbf{\chi}  =  j!k!(n-k-m-j)!m!\,\Xi_{g_\zeta}(j+k)\Xi_{g_\zeta}
(n-k-j),
\end{equation}
whence it follows that
\begin{eqnarray} \lefteqn{
\langle \phi_{n,km}(\alpha_1,\alpha_2,\beta_1,\beta_2)|\phi_{n,km}
(\alpha_1,\alpha_2,\beta_1,\beta_2)\rangle}\quad\nonumber \\ & =&
\frac{|\mathcal{N}(g_\zeta,n)|^2n!^2}{k!m!}\,|u|^{2(n-k-m)}(1-|u|^2)^{k+m}
\nonumber \\ & &{}\times\sum_{j=0}^{n-k-m}\,
\frac{\Xi_{g_\zeta}(j+k)\Xi_{g_\zeta}(n-k-j)}{j!(n-k-m-j)!}\,
\alpha_1^{j+k}\alpha_2^{n-k-j}\beta_1^{n-k-m-j}\beta_2^j \nonumber \\
& = & \mathcal{G}_{n,km}(w,x,y,z),
\end{eqnarray}
where we define
\begin{eqnarray}
\lefteqn{
\mathcal{G}_{n,km}(w,x,y,z)} \quad \nonumber \\ &\defeq &
 \frac{|\mathcal{N}(g_\zeta,n)|^2n!^2}{k!m!} \,\sum_{j=0}^{n-k-m}\,
\frac{\Xi_{g_\zeta}(j+k)\Xi_{g_\zeta}(n-k-j)}{j!(n-k-m-j)!}\,
w^jx^ky^{n-k-m-j}z^m,\qquad
\label{eq:CalGDef}
\end{eqnarray}
with
\begin{equation}
w = \alpha_1\beta_2|u|^2,\quad  x = \alpha_1(1-|u|^2),\quad y=
\alpha_2\beta_1|u|^2,\quad z = \alpha_2(1-|u|^2).
\label{eq:wxyzDef}
\end{equation}
Thus for $\mathbf{L} \subset \{a_1,a_2,b_1,b_2\}$, the recipe of
Sec.\ \ref{sec:8} yields for the $\mathcal{F}_{n,km}$ of Eq.\
(\ref{eq:formM0ch10})
\begin{equation}
\mathcal{F}_{n,km}(\mathbf{L}) \defeq  (-1)^{\#(\mathbf{L})}\left(\prod_{x
\in \mathbf{L}}[1-p_{\rm dark}(x)]\right)
\mathcal{G}_{n,km}(w,x,y,z)
\left|_{\text{Eval}}\right.,
\label{eq:calFDef}
\end{equation}
with $\mathcal{G}_{n,km}(w,x,y,z)$
evaluated, using Eq.\ (\ref{eq:wxyzDef}), according to:
\begin{eqnarray}
\alpha_j & = & \left\{\begin{array}{l}1-\eta_{\rm det}(a_j),\quad \mbox{
if }a_j \in \mathbf{L}, \\ 1,\quad \mbox{ otherwise,} \end{array}
\right. \nonumber \\ \beta_j & = & \left\{\begin{array}{l}1-\eta_{\rm
det}(b_j)\eta_{\rm trans},\quad \mbox{ if }b_j \in \mathbf{L}, \\ 1,\quad
\mbox{ otherwise.} \end{array} \right. 
\label{eq:Feval}
\end{eqnarray}

The remaining probabilities that we need to evaluate are less
fine-grained; they are
\begin{equation}
\langle\psi_n|\mathbf{M}(\overline{a}_1,\overline{b}_2;
  a_2,b_1)|\psi_n\rangle = \sum_{{\scriptstyle k,m = 0}\atop
{\scriptstyle k+m \le n}}
\langle\psi_{n,km}|\mathbf{M}(\overline{a}_1,\overline{b}_2;
  a_2,b_1) |\psi_{n,km} \rangle,
\end{equation} along with
$\langle\psi_n|\mathbf{M}(\overline{a}_2,\overline{b}_1;
  a_1,b_2)|\psi_n\rangle$ and the two error outcomes
  $\langle\psi_n|\mathbf{M}(\overline{a}_2,\overline{b}_2;
  a_1,b_1)|\psi_n\rangle$ and
  $\langle\psi_n|\mathbf{M}(\overline{a}_1,\overline{b}_1;
  a_2,b_2)|\psi_n\rangle$.  These are calculated by replacing
$\mathcal{T}_{n,km}$ by 
\begin{equation}
\mathcal{T}_n(\mathbf{J}_0,\mathbf{J}_1) \defeq  
\sum_{{\scriptstyle k,m = 0}\atop {\scriptstyle k+m \le n}}
\mathcal{T}_{n,km}(\mathbf{J}_0,\mathbf{J}_1),
\end{equation}
the calculation of which is streamlined by noticing, in analogy to
Eq.\ (\ref{eq:calTDef}),
\begin{equation}\mathcal{T}_n(\mathbf{J}_0,\mathbf{J}_1)
= \sum_{\mathbf{X} \subset
\mathbf{J}_1}(-1)^{\#(\mathbf{X})}
\mathcal{F}_n(\mathbf{J}_0\|\mathbf{X}),
\label{eq:calTEval}
\end{equation}
with 
\begin{equation}
\mathcal{F}_n(\mathbf{L}) \defeq  \sum_{{\scriptstyle k,m =
0}\atop {\scriptstyle k+m \le n}}
\mathcal{F}_{n,km}(\mathbf{L}),
\end{equation}
evaluated using 
\begin{equation}
\mathcal{G}_n(w,x,y,z)
\defeq \sum_{{\scriptstyle k,m = 0}\atop
{\scriptstyle k+m \le n}}
\mathcal{G}_{n,km}(w,x,y,z).
\end{equation}
Letting $r = j+k$ and re-ordering the sum and using Eq.\
(\ref{eq:CalGDef}), one finds  
\begin{eqnarray}\lefteqn{
\mathcal{G}_n(w,x,y,z)}\quad \nonumber \\ & = &
|\mathcal{N}(g_\zeta,n)|^2n!^2\sum_{r=0}^n
\Xi_{g_\zeta}(r)\Xi_{g_\zeta}(n-r)  \sum_{j=0}^r
\frac{w^jx^{r-j}}{j!(r-j)!}\sum_{\ell = 0}^{n-r}\frac{y^\ell
  z^{n-r-\ell}}{\ell!(n-r-\ell)!} \nonumber \\
& = &|\mathcal{N}(g_\zeta,n)|^2n!^2\sum_{r=0}^n
\frac{\Xi_{g_\zeta}(r)\Xi_{g_\zeta}(n-r)}{r!(n-r)!}(w+x)^r(y+z)^{n-r}.
\qquad
\label{eq:GnEval}
\end{eqnarray}

Turning to the probabilities that require summing over a distribution
of energies, we use Eq.\ (\ref{eq:mainEval}) to make Eq.\
(\ref{eq:probEn}) explicit:
\begin{equation}
\Pr(\mathbf{J}_0,\mathbf{J}_1,k,m) = 
\mathcal{T}_{\mathbf{C},km}\defeq 
\sum_{n=k+m}^\infty|C_n|^2\mathcal{T}_{n,km},
\label{eq:probEn2}
\end{equation}
where we recognize that $\mathcal{F}_{n,km}$ and $\mathcal{T}_{n,km}$
are zero if $n < k+m$.  Similarly Eq.\ (\ref{eq:probEnn}) becomes
\begin{equation}
\Pr(\mathbf{J}_0,\mathbf{J}_1) =
\mathcal{T}_{\mathbf{C}}(\mathbf{J}_0,\mathbf{J}_1)\defeq 
\sum_{n=k+m}^\infty|C_n|^2\mathcal{T}_{n}(\mathbf{J}_0,\mathbf{J}_1).
\label{eq:probEnn2}
\end{equation}

Given an energy profile $\mathbf{C} \defeq \{|C_n|^2\}$,
one evaluates Eq.\ (\ref{eq:probEn}) efficiently by defining
\begin{equation}
\mathcal{G}_{\mathbf{C},km}(w,x,y,z) =  \sum_{n = k+m}^\infty|C_n|^2
\mathcal{G}_{n,km}(w,x,y,z),\label{eq:calGCkm}
\end{equation} 
and observing that, analogous to Eq.\ (\ref{eq:calFDef}), 
\begin{equation}\mathcal{F}_{\mathbf{C},km}(\mathbf{L}) = 
(-1)^{\#(\mathbf{L})}\left(\prod_{x
\in \mathbf{L}}[1-p_{\rm dark}(x)]\right)
\mathcal{G}_{\mathbf{C},km}(w,x,y,z)
\left|_{\text{Eval}}\right.,
\end{equation}
evaluated by the prescription of Eq.\ (\ref{eq:Feval}).

For evaluating $\Pr(\mathbf{J}_0,\mathbf{J}_1)$ (for use when
one is indifferent to
Evangeline's outcome components), we introduce the analogous
\begin{equation}\mathcal{F}_{\mathbf{C}}(\mathbf{L}) = 
(-1)^{\#(\mathbf{L})}\left(\prod_{x
\in \mathbf{L}}[1-p_{\rm dark}(x)]\right)
\mathcal{G}_{\mathbf{C}}(w,x,y,z)
\left|_{\text{Eval}}\right.,\label{eq:calFCofL}
\end{equation}
evaluated by the prescription of Eq.\ (\ref{eq:Feval}), but with
\begin{equation}
\mathcal{G}_{\mathbf{C}}(w,x,y,z) =  \sum_{n = k+m}^\infty|C_n|^2
\mathcal{G}_{n}(w,x,y,z).
\end{equation}

\subsection{Case I:\ \ No frequency entanglement}

The absence of frequency entanglement is exemplified by $g_{\mathrm
I}(\omega,\tilde{\omega}) = f(\omega)h(\tilde{\omega})$, normalized so
that
$\int\!\!\int d\omega\,d\tilde{\omega}\,|g(\omega,\tilde{\omega})|^2 =
\int d\omega\,|f(\omega)|^2 =
\int d\tilde{\omega}\,|h(\tilde{\omega})|^2 = 1$.  Denoting
$\mathcal{N}(g_\zeta,n)$ evaluated at $\zeta = 0$ by
$\mathcal{N}_{\mathrm I}(n)$, we have from the rules for Case I at the end of
Appendix~\ref{app:C} applied to Eq.\ (\ref{eq:cal2N})
\begin{equation}
|\mathcal{N}_{\mathrm I}(n)|^2 =
\left(n! \sum_{k=0}^n\left(\begin{array}{c}n\\[-5pt] k\end{array}
\right)k!(n-k)!\right)^{-1} =
\frac{1}{(n+1)n!^2}.
\label{eq:NIeval}
\end{equation}
{}From Lemma (\ref{eq:XiI}) of Appendix~\ref{app:C}, we have
\begin{equation}
\Xi_{\mathrm I}(n) \defeq \frac{1}{n!}\,\langle
0|\Bigl(a\ \cdot\stackrel{\textstyle g^*_{\mathrm
I}}{\longrightarrow}\cdot\ b\Bigr)^n \Bigl(a^\dag\
\cdot\stackrel{\textstyle g_{\mathrm I}}{\longrightarrow}\cdot\
b^\dag\Bigr)^n |0\rangle = n!\,.
\label{eq:XiI2}
\end{equation}
With these specializations, Eq.\ (\ref{eq:CalGDef}) becomes
\begin{eqnarray} \lefteqn{
\mathcal{G}^{({\mathrm I})}_{n,km}(w,x,y,z) }\quad
  \nonumber \\ &= & \frac{1}{(n+1)k!m!}\,x^kz^m
  \sum_{j=0}^{n-k-m} \frac{(j+k)!(n-k-j)!}{j!(n-k-m-j)!}\,
  w^jy^{n-k-m-j}.
\label{eq:CalGevalI}
\end{eqnarray}
For reference, we note that this involves a hypergeometric function
\cite{batemanI}
\begin{eqnarray} \lefteqn{
\mathcal{G}^{({\mathrm I})}_{n,km}(w,x,y,z) }\quad 
  \nonumber \\ &=& 
  \frac{(n-k)!}{(n+1)m!(n-k-m)!}\,x^ky^{n-k-m}z^m\,
  _2F_1\left(k+1,k+m-n;k-n;\frac{w}{y}\right).\qquad\quad
 \label{eq:CalGevalIA}
\end{eqnarray}
For use when one is indifferent to Evangeline's outcome components,
one finds 
$\mathcal{G}^{({\mathrm I})}_n(w,x,y,z)$ analytically
from Eq.\ (\ref{eq:GnEval}) as
\begin{eqnarray}
\mathcal{G}^{({\mathrm I})}_n(w,x,y,z) & \defeq  & \sum_{{\scriptstyle k,m = 0}\atop {\scriptstyle k+m \le
n}}\mathcal{G}^{({\mathrm I})}_{n,km}(w,x,y,z)
  \nonumber \\ & = &
  \frac{(y+z)^{n+1}-(w+x)^{n+1}}{(n+1)(y+z-w-x)}.
\label{eq:GIn}
\end{eqnarray}
In numerical work, we encounter the limit as $y+z-w-x
\rightarrow 0$, in which case this becomes
\begin{equation}
\lim_{y+z-w-x \rightarrow 0}\mathcal{G}_n^{({\mathrm I})}(w,x,y,z) =
(y+z)^n.
\end{equation}

For the special case of the Poisson energy distribution, we
sum Eq.\ (\ref{eq:GIn}) to obtain
\begin{eqnarray}
\mathcal{G}_\mu^{({\mathrm I})}(w,x,y,z) & \defeq  &
e^{-\mu}\sum_{n=0}^\infty \frac{\mu^n}{n!} \,\mathcal{G}_n^{({\mathrm
I})}(w,x,y,z)
\nonumber \\ & = &
\frac{e^{-\mu}}{y+z-w-x}\,\frac{1}{\mu}\,\left[e^{\mu(y+z)}-e^{\mu(w+x)}
\right].
\label{eq:GmuI}
\end{eqnarray} 
In numerical work, we again encounter the limit as $y+z-w-x
\rightarrow 0$, in which case this becomes
\begin{equation}
\lim_{y+z-w-x \rightarrow 0}\mathcal{G}_\mu^{({\mathrm I})}(w,x,y,z) =
e^{-\mu(1-y-z)}.
\end{equation}

\subsection{Case II:\ \ Limit of extreme frequency entanglement as
$\zeta \rightarrow \pm \infty$} 

It makes no sense to ask for the limit of $g_\zeta$; however, we
explore how the probabilities behave in the limit of large values of
$|\zeta|$.  Denoting $\mathcal{N}(g_\zeta,n)$ in the limit as $\zeta
\rightarrow \pm \infty$ by $\mathcal{N}_{\mathrm{II}}(n)$, we have from the
rules for Case II at the end of Appendix~\ref{app:C} applied to Eq.\
(\ref{eq:cal2N})
\begin{equation}
|\mathcal{N}_{\mathrm{II}}(n)|^2 = \frac{2^{-n}}{n!},
\label{NIIeval}
\end{equation}
which (in its power of $n!$) differs from the preceding case.
{}From Lemma (\ref{eq:XiII}) of Appendix~\ref{app:C} we have
\begin{equation}
\Xi_{\mathrm{II}}(n) \defeq 
\lim_{\zeta \rightarrow \pm \infty}
\frac{1}{n!}\,\langle 0|\Bigl(a\ \cdot\stackrel{\textstyle
g^*}{\rightarrow}\cdot\ b\Bigr)^n \Bigl(a^\dag\ \cdot\stackrel{\textstyle
g}{\rightarrow}\cdot\ b^\dag\Bigr)^n|0\rangle = 1. 
\label{eq:XiIIcopy}
\end{equation}
With these specializations, Eq.\ (\ref{eq:CalGDef}) becomes, for Case II,
\begin{eqnarray} \lefteqn{
\mathcal{G}^{(\mathrm{II})}_{n,km}(w,x,y,z) }\quad
  \nonumber \\ &=& \frac{2^{-n}n!}{k!m!}
  \sum_{j=0}^{n-k-m} \frac{1}{j!(n-k-m-j)!}
  w^jx^ky^{n-k-m-j}z^m \nonumber
  \\ & =&
  \frac{2^{-n}n!}{k!m!(n-k-m)!}\,x^kz^m
  (w+y)^{n-k-m}.
\label{eq:CalGevalII}
\end{eqnarray}
Putting this together with the Poisson distribution for energy
yields
\begin{eqnarray}
\mathcal{G}^{(\mathrm{II})}_{\mu,km}(w,x,y,z) &
\defeq  & \sum_{n = k+m}^\infty |C_n|^2
\mathcal{G}^{(\mathrm{II})}_{n,km}(w,x,y,z) \nonumber
\\ & = & e^{-\mu}\sum_{n=k+m}^\infty
\left(\frac{\mu}{2}\right)^n\,\frac{x^kz^m(w+y)^{n-k-m}}{k!m!(n-k-m)!}
\nonumber
\\ & = &
\exp\left[-\frac{\mu}{2}(2-w-y)\right]\left(\frac{\mu}{2}\right)^{k+m}\,
\frac{x^kz^m}{k!m!}.
\end{eqnarray}

For the sum of these over $k,m$, Eqs.\ (\ref{eq:XiIIcopy}) and
(\ref{eq:GnEval}) imply
\begin{equation}
\mathcal{G}^{(\mathrm{II})}_n(w,x,y,z) =
2^{-n}(w+x+y+z)^n.
\end{equation}
Summing over all $n$ weighted by $|C_n|^2$ yields, for use in
calculating probabilities for Alice and Bob regardless of Evangeline's
outcome components,
\begin{eqnarray}
\mathcal{G}^{(\mathrm{II})}_\mu(w,x,y,z) &
\defeq  &  e^{-\mu}\sum_{n=k+m}^\infty \frac{\mu^n}{n!}\,
2^{-n}(w+x+y+z)^n
\nonumber \\ & = & \exp{\left[-\frac{\mu}{2}(2 -w -x-y-z)\right]} .
\end{eqnarray}

\subsection{Example numbers}\label{subsec:10G}

\begin{figure}[t]  
\begin{center}
\includegraphics[width=5.5in]{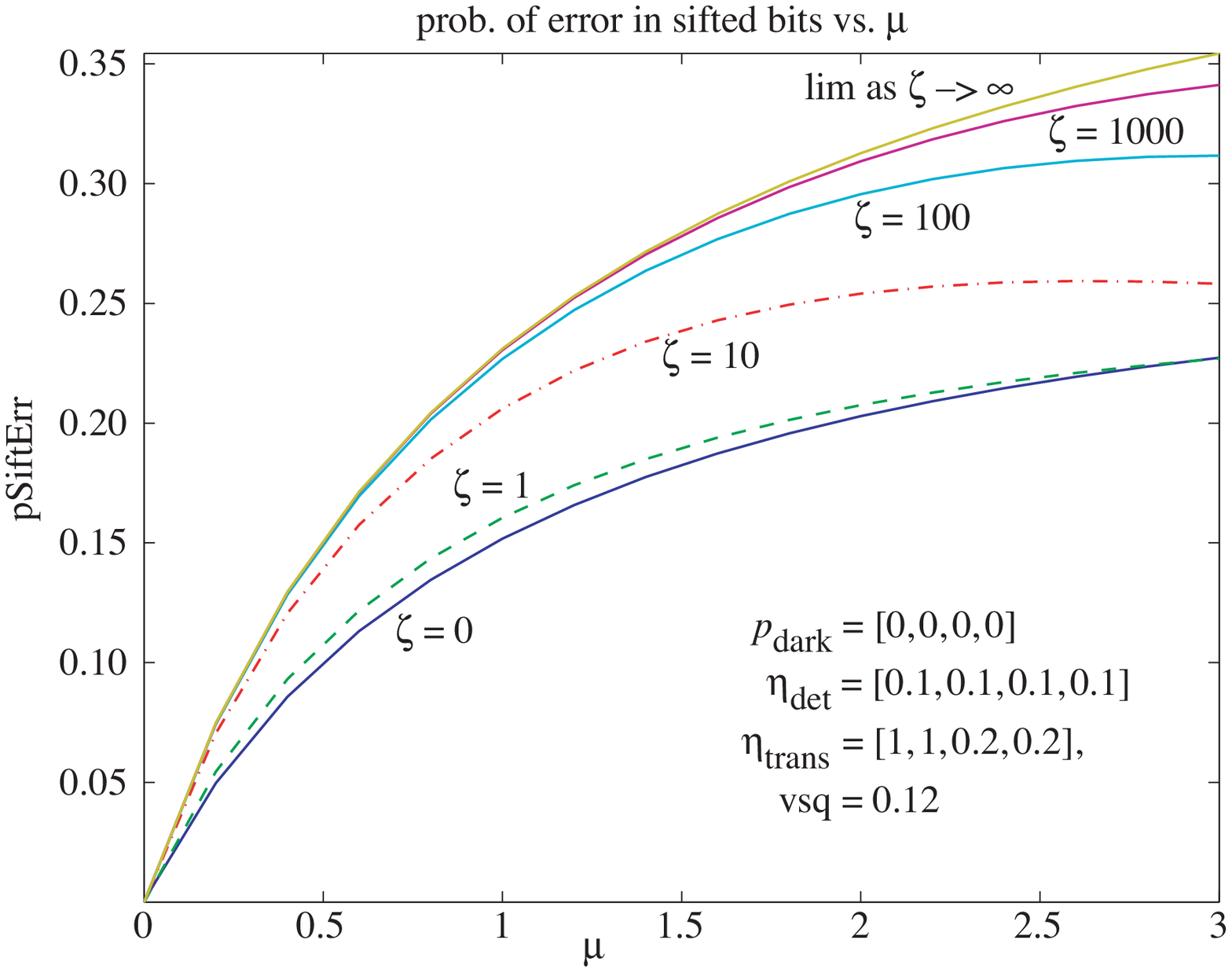}
\end{center}
\vspace{-24pt}
\caption{Probability of error versus $\mu$, for $\zeta=0$, 1, 10,
100, 1000, and $\zeta\to\infty$.}\label{fig:7}
\vspace{40pt}\end{figure}

Figure \ref{fig:7} shows the probability of error vs.\ $\mu$ for several
values of $\zeta$.  This shows explicitly how changing frequency
entanglement $\zeta$ changes the dependence of the probability of error on
$\mu$. Other cases can be generated from the MATLAB programs listed in
Appendix~\ref{app:F}.

\appendix
\section{Background}\label{app:A}

\subsection{Commutation relation}
The commutation relation was chosen by analogy with that for the
harmonic oscillator.  One can ask if the $\delta$-function should
be multiplied by a factor that depends on propagation constant.
We answer ``no'' for the following reason.  We want the energy
of a 1-photon light state having a narrow frequency spectrum
centered at $\omega_0$ to be close to $\hbar |\omega_0|$.  Taking
such a state as an in-state to a fiber-vacuum interface results
in an out-state of the same frequency but different wavelength.
In order for energy to be conserved, we need the energy calculated
for a 1-photon state to be independent of variations in wavelength
at a given frequency.  That rules out any factor in the commutation
relation that depends on the in-fiber propagation constant.

\subsection{Units} $a(\omega)$ in units of (seconds)$^{1/2}$;
$a_f$ is dimensionless for a normalized function $f$ which
has dimension of sec$^{1/2}$. 
  
Viewing a single-mode of a path as a transmission line, we need
an operator corresponding to voltage (analogous to the electric-field
operator of quantum electrodynamics).

\subsection{Quantum mechanics stripped of space and time}
 
Often it is convenient to make a preliminary analysis that skips
all the integrals over frequency by treating quantum states in 
a toy Hilbert space of finite dimension, which means that space and
time are collapsed to zero dimensions.  (That still leaves
polarization, for example.)  This procedure is equivalent to
an analysis allowing for frequency for certain calculations, namely
when the frequency functions involved are all mutually orthogonal.
An example is Eq.\ (\ref{eq:coh}).  

\section{Operator lemmas}\label{app:B}

{}For any two operators $A$ and $B$ let $[A,B] \defeq  AB
-BA$. 

\vspace{3pt}

\noindent\textbf{Lemma}: For any operators $A$, $B$, $C$,
\begin{eqnarray} [A,BC] & = & [A,B]C + B[A,C], \nonumber \\  
{}[AB,C] & = & [A,C]B + A[B,C].
\label{eq:lem1}\end{eqnarray}

\vspace{3pt}

\noindent\textbf{Lemma}: If $[B,C] = 0$, then 
\begin{eqnarray}
[[A,B],C] & = & [[A,C],B], \nonumber \\
{}[C,[B,A]] & = & [B,[C,A]].
\label{eq:swap}
\end{eqnarray}

\vspace{3pt}

\noindent\textbf{Lemma}: For any four operators $A$, $B$, $C$, $D$,
\begin{equation}[AB,CD] = A[B,C]D + [A,C]BD+ C[A,D]B + CA[B,D].
\label{eq:lem2a}
\end{equation}\vspace{3pt}

\noindent\textbf{Lemma}: For $n = 1$, $2$, \dots,
\begin{equation}[A,B^n] = \sum_{k=1}^nB^{k-1}[A,B]B^{n-k}.
\label{eq:lemB2}
\end{equation}

\vspace{3pt}

\noindent\textbf{Lemma}: If $A|0\rangle = B|0\rangle = 0$, then
\begin{equation}
\langle 0|AB A^\dag B^\dag|0\rangle =
\langle 0|A[B,A^\dag]B^\dag|0\rangle
+\langle 0|[A,A^\dag][B,B^\dag]|0\rangle.
\end{equation}

\noindent\textbf{Lemma}: For any two operators $A$ and $B^\dag$,
 if $[[[A,B^\dag],B^\dag],B^\dag] = 0$, then
\begin{equation}
[A,B^{\dag n}]= n\left\{\frac{n-1}{2}B^{\dag(n-2)}[[A,B^\dag],B^\dag] +
B^{\dag(n-1)}[A,B^\dag]\right\}.
\label{eq:lem33}
\end{equation}

\noindent\textit{Proof}:\quad  $[A,B^{\dag n}]  =  \sum_{k=1}^n B^{\dag
  (k-1)}[A,B^\dag]B^{\dag(n-k)}$ and 
\begin{eqnarray}[A,B^\dag]B^{\dag(n-k)} & = &
B^{\dag (n-k)}[A,B^\dag]+[[A,B^\dag],B^{\dag(n-k)}] \nonumber \\ & = &
B^{\dag (n-k)}[A,B^\dag]+
\sum_{j=1}^{n-k}B^{\dag(n-k-1)}[[A,B^\dag],B^\dag], \nonumber
\end{eqnarray}
whence the lemma follows. $\Box$

\vspace{3pt}

\noindent\textbf{Lemma}: If $[[[A,A^\dag],A^\dag],A^\dag] = 0$ and
$A|0\rangle = 0$, then
\begin{equation} \langle 0|A^n A^{\dag
n}|0\rangle = \frac{n(n-1)}{2}\langle
0|A^{n-1}A^{\dag(n-2)}[[A,A^\dag],A^\dag]|0\rangle +
n\langle 0|AA^\dag|0\rangle \langle 0|A^{n-1} A^{\dag (n-1)}|0\rangle.
\label{eq:AnAn}
\end{equation}
\noindent\textit{Proof}: Notice that $\langle 0|A^n A^{\dag
n}|0\rangle = \langle 0|A^{n-1}[A,A^{\dag
n}]|0\rangle$ and use Lemma (\ref{eq:lem33}). $\Box$

Lemma (\ref{eq:AnAn}) shows how repeated commutators work their way
into $\langle 0|A^n A^{\dag n}|0\rangle$.

\vspace{3pt}

\noindent\textbf{Lemma}: Given $A|0\rangle = B|0\rangle = 0$ and
$[[A,B^\dag],B^\dag] = [A,[A,B^\dag]] = 0$, then
\begin{equation}
\langle 0|A^nB^{\dag n}|0 \rangle = n \langle 0|A^{n-1} B^{\dag
(n-1)}[A,B^\dag]|0\rangle,
\end{equation} from which it follows that:

\vspace{3pt}

\noindent\textbf{Lemma}: Given $A|0\rangle = B|0\rangle = 0$ and
$[[A,B^\dag],B^\dag] = [A,[A,B^\dag]] = 0$, then
\begin{equation}
\langle 0|[A^n,B^{\dag n}]|0\rangle = n!\langle 0|[A,B^\dag]^n|
0\rangle. 
\label{eq:lemComII} 
\end{equation} 
Note: if $A = a_f$, then $n!^{-1/2}A^{\dag n}|0\rangle$ is an $n$-photon
state; thus two $n$-photon states of this type have as their inner
product the $n$-th power of the inner product of the corresponding
1-photon states.  It follows that a unitary transform can
convert an $n$-photon state into a tensor product of an $(n-1)$-photon
state and a 1-photon state.
\vspace{3pt}

\noindent\textbf{Lemma}: For any two operators $A$ and $B$ such that
$[B,[A,B]] = 0$ and $n = 0$, $1$, $2$, \dots, we have 
\begin{equation}[A,B^n] =
n[A,B]B^{n-1}.
\end{equation}

(Proof follows by induction, using Lemma (\ref{eq:lem1}).)

\vspace{3pt}

\noindent\textbf{Lemma}: For any two operators $A$ and $B$ such that
$[B,[A,B]] = 0$, we have 
\begin{equation}[A,\exp(B)] = [A,B]\exp(B).
\end{equation}

(Proof by expansion of exponential, using Lemma (\ref{eq:lem2a}).)

\vspace{3pt}

\noindent\textbf{Lemma}: Given any operators $A$, $B_1,\ldots, B_n$,
\begin{equation}
\left[A,\prod_{j=1}^n B_j\right] = \sum_{\ell=1}^n
  \left(\prod_{j=1}^{\ell-1}
  B_j\right)[A,B_\ell]\left(\prod_{j=\ell+1}^n B_j\right),
\label{eq:commProd}
\end{equation}
with the convention that for any $m > n$ and any $X_j$
\begin{equation}\prod_{j=m}^n X_j = 1.
\label{eq:conven}
\end{equation}

\subsection{Implications of commutation rules}
Assume for the rest of this appendix the commutation rules
Eqs.\ (\ref{eq:comm0}) and (\ref{eq:comm1}).  Then we have

\vspace{3pt}

\noindent\textbf{Lemma}: For a set of frequencies
$\omega, \omega_1, \ldots, \omega_n$, with $n \ge 1$,
\begin{equation} \left[a(\omega), \prod_{j=1}^n a^\dag(\omega_j)\right]
= \sum^n_{j=1} \delta(\omega - \omega_j)
\prod_{{\scriptstyle k=1}\atop{\scriptstyle \, k \ne j}}^n
a^\dag(\omega_k).
\label{eq:commsing}
\end{equation}   

\vspace{3pt}

\noindent\textbf{Lemma}:  Let $S_n$ be the permutation group on
$1,\ldots,n$; then
\begin{equation}
\langle 0|\prod_{j=1}^m a(\omega'_j)\prod_{k=1}^n
a^\dag(\omega_k)|0\rangle = \delta_{mn}\sum_{\pi \in
  S_n}\prod_{j=1}^n\delta(\omega_j - \omega'_{\pi j}).
\label{eq:commnn}
\end{equation}
{}From this follows

\vspace{3pt}

\noindent \textbf{Lemma}: 
\begin{equation}
\langle 0|(g^*_m\!:\!a^m)(h_n\!:\!a^{\dag n})|0\rangle =\delta_{mn}
n!\int \!d\omega_1 \cdots d\omega_n\,
g^*_m(\omega_1,\ldots,\omega_n)\mathcal{S}(\omega_1,\ldots,\omega_n)
h_n(\omega_1,\ldots,\omega_n),
\end{equation}
where $\mathcal{S}$ is defined in (\ref{eq:Sdef}).  From this follows
another useful fact of norms:

\vspace{3pt}

\noindent\textbf{Lemma}: If $h_n(\omega_1,\ldots,\omega_n)$ is
symmetric under all permutations of its arguments, then
\begin{equation}
\langle 0|(h^*_n\!:\!a^n)(h_n\!:\!a^{\dag n})|0\rangle = n!\int
d\omega_1\cdots d\omega_n\,|h_n(\omega_1,\ldots,\omega_n)|^2.
\label{eq:normhn}
\end{equation}
We also have the following relation that allows the calculation
of some detection probabilities:

\vspace{3pt}

\noindent\textbf{Lemma}: If $[b(\omega),a^\dag(\omega')]=0$ and
$h_n(\omega_1,\ldots,\omega_n)$ is symmetric under all permutations
of its arguments, then 
\begin{equation}
\langle 0|(h^*_n\!:\!a^kb^{n-k})(h_n\!:\!a^{\dag k}
b^{\dag(n-k)})|0\rangle
=\frac{k!(n-k)!}{n!}\,\langle 0|(h^*_n\!:\!a^n)(h_n\!:\!a^{\dag n})|0\rangle.
\label{eq:nkrelation}
\end{equation}

\vspace{5pt}

\noindent\textit{Proof}:
\begin{eqnarray} \lefteqn{
\langle 0|(h^*_n\!:\!a^kb^{n-k})(h_n\!:\!a^{\dag k}
b^{\dag(n-k)})|0\rangle  }\quad \nonumber \\ &=& \int d\omega_1\cdots
d\omega_n\,d\omega'_1\cdots d\omega'_n\,
h_n^*(\omega_1,\ldots,\omega_n)h_n(\omega'_1,\ldots,\omega'_n)
\nonumber \\ & & \langle 0|a(\omega_1)\cdots
a(\omega_k)b(\omega_{k+1})\cdots b(\omega_n)a^\dag(\omega'_1)\cdots
a^\dag(\omega'_k)b^\dag(\omega'_{k+1})\cdots b^\dag(\omega'_n)|0
\rangle. \nonumber
\end{eqnarray}
Because $a$ and $b$ work on distinct tensor-product factors, we have
\begin{eqnarray} \lefteqn{
\langle 0|a(\omega_1)\cdots a(\omega_k)b(\omega_{k+1})\cdots
b(\omega_n)a^\dag(\omega'_1)\cdots
a^\dag(\omega'_k)b^\dag(\omega'_{k+1})\cdots b^\dag(\omega'_n)|0
\rangle }\quad \nonumber \\ & = & \langle 0|a(\omega_1)\cdots
a(\omega_k)a^\dag(\omega'_1)\cdots a^\dag(\omega'_k)|0\rangle \langle
0|b(\omega_{k+1})\cdots b(\omega_n)b^\dag(\omega'_{k+1})\cdots
b^\dag(\omega'_n)|0 \rangle. \nonumber
\end{eqnarray}
The lemma then follows from the symmetry of $h_n$ together with Lemma
(\ref{eq:commnn}). $\Box$

Concerning broad-band coherent states, from Lemma (\ref{eq:commsing})
follows:

\vspace{3pt}

\noindent\textbf{Lemma}:  For $n \ge 1$, 
\begin{equation}
a(\omega) (a_f^\dag)^n|0\rangle = n f(\omega) (a_f^\dag)^{n-1}|0\rangle.
\label{eq:lem6}\end{equation}

\vspace{3pt}

\noindent\textbf{Lemma}:  For the coherent
state defined by Eq.\ (\ref{eq:coh}),
\begin{equation} 
a_g|\alpha,a_f\rangle = \int d\omega\, g^*(\omega)a(\omega) |\alpha,
a_f\rangle = \alpha\left(\int d\omega\,
g^*(\omega)f(\omega)\right)|\alpha, a_f\rangle.
\end{equation}

\vspace{3pt}

\noindent\textbf{Lemma}: For $P_n(a_1,a_2)$ defined in
Eq.\ (\ref{eq:Pdist}), 
\begin{equation} \sum_{n=0}^\infty n P_n(a_1,a_2) = \int d\omega
\sum_{j=1}^2a_j^\dag(\omega)a_j(\omega).
\label{eq:b7}
\end{equation}

\vspace{3pt}

\noindent\textit{Proof}:\quad By the definition of Eq.\ (\ref{eq:Pdist}),
we have 
\begin{eqnarray}
\sum_{n=0}^\infty n P_n(a_1,a_2) & = & \sum_{n=0}^\infty\ \sum_{k=0}^n
P_k(a_1)P_{n-k}(a_2) \nonumber \\ & = &\sum_{k=0}^\infty \left(k
P_k(a_1) \sum_{n=k}^\infty P_{n-k}(a_2) + P_k(a_1)\sum_{n=k}^\infty
(n-k)P_{n-k}(a_2)\right).\qquad
\end{eqnarray} 
The lemma then follows from Eqs.\ (\ref{eq:unit}) and
(\ref{eq:number}).

\section{Algebra of frequency-entangled operators}\label{app:C}
We want to evaluate expressions of the form $\langle
0|\mbox{Pol}^\dag\mbox{Pol}|0\rangle$, where $\mbox{Pol}$ is a
polynomial in annihilation operators.  The general method of
evaluation is to use commutation relations to rearrange the operators
so that, in the end, nothing is left but a number.  The commutation
relations amount to an algebra, which we now construct for the
simplest quantum models that show how polarization entanglement
combines with frequency entanglement. The models cover bi-SU(2)
invariant states built up from polynomials in operators of the form
\hbox{$g\!:\!a_j^\dag b_k^\dag$}.  Consider some number $N_a$ of modes
derived from $a(\omega)$ and $N_b$ of modes derived from $b(\omega)$:
\begin{equation} a_j(\omega), b_k(\omega), \quad\mbox{ for } j = 1,\dots,
    N_a, \mbox{ and\ } k = 1, \dots, N_b,
\label{eq:modes}
\end{equation}  
with 
\begin{equation} a_j(\omega)|0\rangle = b_k(\omega)|0\rangle = \langle 0|
    a_j^\dag(\omega) = \langle 0| b_k^\dag(\omega) = 0 .
\label{eq:kill2}
\end{equation}
The commutation relations are
\begin{eqnarray}
{}[a_j(\omega),a_k^\dag(\omega')]& = & \delta_{jk}\delta(\omega-\omega'),
\nonumber \\ 
{}[b_j(\omega),b_k^\dag(\omega')]& = & \delta_{jk}\delta(\omega-\omega'),
\nonumber \\
{}[a_j(\omega),b_k^\dag(\omega')]& = & 
{}[a_j(\omega),a_k(\omega')]= [b_j(\omega),b_k(\omega')] = 0.
\label{eq:commjk}
\end{eqnarray}

\subsection{Arrow notation for frequency dependence}

Let $p$ and $q$ denote any of these improper creation or annihilation
operators.  We consider operators of the form
\begin{equation}
p\ \cdot\stackrel{\textstyle  g}{\rightarrow}\cdot\ q \defeq 
\int \! \int \!d\omega\,d\tilde{\omega} \, p(\omega)
g(\omega,\tilde{\omega})q(\tilde{\omega})
\label{eq:opdef}.
\end{equation}
Let $a^\dag$ and $b^\dag$ denote any of the improper creation
operators and fix a square-integral function
$g(\omega,\tilde{\omega})$; we are interested in the commutator
algebra generated by
\begin{equation} 
a^\dag\ \cdot\stackrel{\textstyle  g}{\rightarrow}\cdot\ b^\dag  
\end{equation} 
and its adjoint, which is
\begin{equation}  (a^\dag\ \cdot\stackrel{\textstyle  g}{\rightarrow}
\cdot\ b^\dag)^\dag = b\ \cdot\stackrel{\textstyle  g^*}{\leftarrow}
\cdot\ a =
  a\ \cdot\stackrel{\textstyle  g^*}{\rightarrow}\cdot\ b.
\end{equation}  

We now develop this arrow notation. For any function of two
variables $g(x_1,x_2)$ that enters as a factor in an integrand, we
write $g$ as $\stackrel{\textstyle  g}{\rightarrow}$ if the second variable is
identified with a variable in a following factor or if the first
variable is identified with a variable in a preceding factor; we write
$\stackrel{\textstyle  g}{\leftarrow}$ if the first variable is identified with a
variable in a following factor or if the second variable is identified
with a variable in a preceding factor. This makes it easy to express
compound convolutions that will occur in commutators, such as
$G_{\mathrm{I}}(x_1,x_2) = \int \! \int \!dx\,dx'\,
f(x_1,x)g(x',x)h(x',x_2)$ and
$G_{\mathrm{II}}(x_1,x_2) = \int \! \int \!dx\,dx'\,
f(x_1,x)g(x,x')h(x',x_2)$; because the order is different in the middle
factor, $G_{\mathrm I}$ and $G_{\mathrm{II}}$ are distinct.  Writing a
``$\cdot$'' for integration, we then diagram
\begin{eqnarray}
G_{\mathrm I}(x_1,x_2) & \mbox{ as } & x_1
\stackrel{\textstyle  f}{\rightarrow} \cdot
\stackrel{\textstyle  g}{\leftarrow} \cdot \stackrel{\textstyle 
h}{\rightarrow} x_2,\nonumber\\ 
G_{\mathrm{II}}(x_1,x_2) & \mbox{ as } & x_1 \stackrel{\textstyle 
f}{\rightarrow} \cdot
\stackrel{\textstyle  g}{\rightarrow} \cdot \stackrel{\textstyle 
h}{\rightarrow} x_2.
\label{eq:arrowdef}
\end{eqnarray}
Altogether there are eight such functions, corresponding to the eight
ways to orient a sequence of three arrows.

A compound function such as $G_{\mathrm I}$ can itself enter another
compound. If $G$ is defined by an arrow diagram with the above procedure,
then $ \stackrel{\textstyle  G}{\rightarrow}$ is obtained immediately as
that diagram while $ \stackrel{\textstyle  G}{\leftarrow}$ is obtained by
left-right reflection of the diagram, as in
\begin{eqnarray}
 \stackrel{\textstyle  G_{\mathrm I}}{\longrightarrow}& = &
\stackrel{\textstyle  f}{\rightarrow}\cdot
\stackrel{\textstyle  g}{\leftarrow} \cdot \stackrel{\textstyle 
h}{\rightarrow},\nonumber \\
 \stackrel{\textstyle  G_{\mathrm I}}{\longleftarrow}& = &
\stackrel{\textstyle  h}{\leftarrow} \cdot
\stackrel{\textstyle  g}{\rightarrow} \cdot \stackrel{\textstyle 
f}{\leftarrow}.
\label{eq:compound}
\end{eqnarray}
We abbreviate repetitive patterns by an exponent; for example
\begin{equation}
\stackrel{\textstyle  g}{\rightarrow} \cdot\stackrel{\textstyle 
h}{\leftarrow}
\cdot\stackrel{\textstyle  g} {\rightarrow}\cdot\stackrel{\textstyle 
h}{\leftarrow}\cdot
\stackrel{\textstyle  g}{\rightarrow}\; =\,
\Bigl(\stackrel{\textstyle  g}{\rightarrow}\cdot \stackrel{\textstyle 
h}{\leftarrow}\Bigr)^2\cdot\stackrel{\textstyle  g}{\rightarrow}.
\end{equation}
We define an operation `Loop' that produces a number from a string
of arrows by joining the two terminal points; for example
\begin{eqnarray}
\mbox{Loop}\Bigl(\stackrel{\textstyle  g}{\rightarrow}\cdot\stackrel{\textstyle  h}
{\leftarrow}\Bigr)&=&(g^*,h) = \int\!\!\int d\omega\,
d\tilde{\omega}\,g(\omega,\tilde{\omega})h(\omega,\tilde{\omega}).
\end{eqnarray}

\noindent\textbf{Lemma}:
\begin{equation}
\stackrel{\textstyle  h}{\leftarrow}\cdot\,
\Bigl(\stackrel{\textstyle  g}{\rightarrow}\cdot\stackrel{\textstyle 
h}{\leftarrow}\Bigr)^n\cdot
\stackrel{\textstyle  g}{\rightarrow}\; = \Bigl(\stackrel{\textstyle 
h}{\leftarrow}\cdot
\stackrel{\textstyle  g}{\rightarrow}\Bigr)^{n+1}.
\label{eq:shift}
\end{equation}

\vspace{3pt}

\noindent\textbf{Lemma}: With these arrow rules, if $ [p,q] = 0$, then
\begin{equation}
 p\ \cdot\stackrel{\textstyle g}{\rightarrow}\cdot\ q =
q\ \cdot\stackrel{\textstyle g}{\leftarrow}\cdot\ p.
\end{equation}

\subsection{Algebra rules} 
Equations (\ref{eq:commjk}) imply a commutator algebra generated by
\begin{equation}
a_j\ \cdot\stackrel{\textstyle g^*}{\rightarrow}\cdot\ b_k \quad 
\text{and}\quad a_j^\dag\ \cdot\stackrel{\textstyle g}{\rightarrow}\cdot\
b_k^\dag.
\label{eq:genjk}
\end{equation}  
The commutator of these, along with commutators of commutators, etc.,
generate new frequency functions.  Regardless of the frequency
function $g$, the operator $a_j\ \cdot\stackrel{\textstyle g}{\rightarrow}
\cdot\ b_k$ acting to the right on the vacuum state annihilates the
state.  We will speak of such an operator, regardless of its frequency
function, as being of {\em annihilation type}. In addition to the creation
and annihilation types that we start with, the commutation relations
engender two more types, one type has the form \hbox{$a_j^\dag\ \cdot
\rightarrow \cdot\ a_k$} or \hbox{$b_j^\dag\ \cdot \rightarrow \cdot\
b_k$}, the other type is just a number.  The point is to evaluate
ex\-pressions of the form $\langle 0|\mbox{Pol}^\dag\mbox{Pol}|0\rangle$
by using the commutator algebra to transform this to a form $\langle
0|x|0\rangle$, where $x$ is just a number, extracted in the last step
from the normalization relation $\langle 0|0\rangle = 1$.

The commutation relations among all these types are defined by the
following and their hermitian conjugates:
\begin{eqnarray}
[a_j\ \cdot\stackrel{\textstyle g}{\rightarrow}\cdot\ b_k,
a_\ell^\dag\ \cdot\stackrel{\textstyle h}{\rightarrow}\cdot\ b_m^\dag] &
= &
\delta_{km}\, a_\ell^\dag\ \cdot\stackrel{\textstyle h}{\rightarrow}
\cdot\stackrel{\textstyle g}{\leftarrow}\cdot\ a_j +\delta_{j\ell}\,
b_m^\dag\ \cdot\stackrel{\textstyle h}{\leftarrow}
\cdot\stackrel{\textstyle g}{\rightarrow}\cdot\ b_k \nonumber \\ & &{} +
\delta_{j\ell}\delta_{km}\mbox{Loop}\Bigl(\stackrel{\textstyle
h}{\leftarrow}
\cdot\stackrel{\textstyle g}{\rightarrow}\Bigr), \\
{}[a_j^\dag\ \cdot\stackrel{\textstyle g}{\rightarrow}\cdot\ a_k,
a_\ell^\dag\ \cdot\stackrel{\textstyle h}{\rightarrow}\cdot\ b_m^\dag] & = &
\delta_{k\ell}\,a_j^\dag\ \cdot\stackrel{\textstyle g}{\rightarrow}
\cdot\stackrel{\textstyle h}{\rightarrow}\cdot\ b_m^\dag, \\
{}[a_j^\dag\ \cdot\stackrel{\textstyle g}{\rightarrow}\cdot\ a_k,
a_\ell^\dag\ \cdot\stackrel{\textstyle h}{\rightarrow}\cdot\ a_m^\dag] & = &
\delta_{k\ell}\,a_j^\dag\ \cdot\stackrel{\textstyle g}{\rightarrow}
\cdot\stackrel{\textstyle h}{\rightarrow}\cdot\ a_m -
\delta_{jm}\,a_\ell^\dag\ \cdot\stackrel{\textstyle h}{\rightarrow}
\cdot\stackrel{\textstyle g}{\rightarrow}\cdot\ a_k, \\
{}[b_j^\dag\ \cdot\stackrel{\textstyle g}{\rightarrow}\cdot\ b_k,
a_\ell^\dag\ \cdot\stackrel{\textstyle h}{\rightarrow}\cdot\ b_m^\dag] & = &
\delta_{km}\,a_\ell^\dag\ \cdot\stackrel{\textstyle h}{\rightarrow}
\cdot\stackrel{\textstyle g}{\leftarrow}\cdot\ b_j^\dag, \\
{}[b_j^\dag\ \cdot\stackrel{\textstyle g}{\rightarrow}\cdot\ b_k,
b_\ell^\dag\ \cdot\stackrel{\textstyle h}{\rightarrow}\cdot\ b_m^\dag] & = &
\delta_{k\ell}\,b_j^\dag\ \cdot\stackrel{\textstyle g}{\rightarrow}
\cdot\stackrel{\textstyle h}{\rightarrow}\cdot\ b_m -
\delta_{jm}\,b_\ell^\dag\ \cdot\stackrel{\textstyle h}{\rightarrow}
\cdot\stackrel{\textstyle g}{\rightarrow}\cdot\ b_k.
\end{eqnarray}
These relations hold for any integrable functions $g(\omega,\omega')$
and $h(\omega,\omega')$.

\subsection{Subalgebra for fixed $g$}
{}For constructing examples, we will use a subalgebra in which a chosen
$g(\omega,\omega')$ plays a distinguished role.  Given any such $g$,
define for $n = 1, 2$, \dots,
\begin{eqnarray}
\stackrel{\textstyle g_n}{\longrightarrow} & = &
\Bigl(\stackrel{\textstyle g}{\rightarrow}
\cdot\stackrel{\textstyle g^*}{\leftarrow}\Bigr)^{n-1}\cdot
\stackrel{\textstyle g}{\rightarrow},
\nonumber \\ \stackrel{\textstyle g_n}{\longleftarrow} & = &
\stackrel{\textstyle g}{\leftarrow} \cdot\,\Bigl(\stackrel{\textstyle
g^*}{\rightarrow}\cdot
\stackrel{\textstyle g}{\leftarrow}\Bigr)^{n-1}, \nonumber \\
\stackrel{\textstyle \check{G}_n}{\longrightarrow} & = &
\Bigl(\stackrel{\textstyle g}{\rightarrow} \cdot\stackrel{\textstyle
g^*}{\leftarrow}\Bigr)^n,
\nonumber \\ \stackrel{\textstyle \widehat{G}_n}{\longrightarrow} & = &
\Bigl(\stackrel{\textstyle g}{\leftarrow} \cdot\stackrel{\textstyle
g^*}{\rightarrow}\Bigr)^n.
\label{eq:gn2Def}
\end{eqnarray}
By writing out a few arrow expressions, one shows for positive
integers $r$ and $s$
\begin{eqnarray}
\stackrel{\textstyle g_s}{\longrightarrow} \cdot\stackrel{\textstyle
g^*_r}{\longleftarrow} & = & \Bigl(\stackrel{\textstyle g}{\rightarrow}
\cdot\stackrel{\textstyle g^*}{\leftarrow}\Bigr)^{r+s-1} =
\stackrel{\textstyle \check{G}_{r+s-1}}{\longrightarrow}, \\
\stackrel{\textstyle g_s}{\longleftarrow} \cdot\stackrel{\textstyle
g^*_r}{\longrightarrow} & = & \Bigl(\stackrel{\textstyle g}{\leftarrow}
\cdot\stackrel{\textstyle g^*}{\rightarrow}\Bigr)^{r+s-1}, \\
\stackrel{\textstyle
\check{G}_r}{\longrightarrow}\cdot\stackrel{\textstyle g_s}
{\longrightarrow} & = & \stackrel{\textstyle g_{r+s}} {\longrightarrow},
\\ \stackrel{\textstyle g_s}{\longrightarrow}\cdot\stackrel{\textstyle
\widehat{G}_r} {\longleftarrow} & = & \stackrel{\textstyle g_{r+s}}
{\longrightarrow}, \\
\stackrel{\textstyle
\check{G}_r}{\longrightarrow}\cdot\stackrel{\textstyle \check{G}_s}
{\longrightarrow} & = &
\stackrel{\textstyle \check{G}_{r+s}}{\longrightarrow},
\\ \stackrel{\textstyle
\widehat{G}_r}{\longrightarrow}\cdot\stackrel{\textstyle \widehat{G}_s}
{\longrightarrow} & = &
\stackrel{\textstyle \widehat{G}_{r+s}}{\longrightarrow}, \\
\stackrel{\textstyle \widehat{G}_n}{\longrightarrow}& = &
\stackrel{\textstyle \widehat{G}^*_n} {\longleftarrow}, \\
\stackrel{\textstyle \check{G}_n}{\longrightarrow}& = &
\stackrel{\textstyle \check{G}^*_n} {\longleftarrow}.
\end{eqnarray} 
With these and the definition in Eq.\ (\ref{eq:gn2Def}), we find
\begin{eqnarray}
\Bigl[a_j\ \cdot\stackrel{\textstyle g^*_r}{\longrightarrow}\cdot\ b_k,\,
a_\ell^\dag\ \cdot\stackrel{\textstyle g_s}{\longrightarrow}\cdot\
b_m^\dag\Bigr] & = &
\delta_{km}\,
a_\ell^\dag\ \cdot\stackrel{\textstyle \check{G}_{r+s-1}}{\longrightarrow}
\cdot\ a_j +\delta_{j\ell}\, b_m^\dag\ \cdot
\stackrel{\textstyle \widehat{G}_{r+s-1}}{\longrightarrow} \cdot\ b_k
\nonumber \\ & &{}+
\delta_{j\ell}\delta_{km}\mbox{Loop}(\widehat{G}_{r+s-1}),\,
\label{eq:comm27}\\
{}\Bigl[a_j^\dag\ \cdot\stackrel{\textstyle \check{G}_r}{\longrightarrow}\cdot\
a_k,\, a_\ell^\dag\ \cdot\stackrel{\textstyle g_s}{\longrightarrow}\cdot\ 
b_m^\dag\Bigr] & = &
\delta_{k\ell}\,a_j^\dag\ \cdot\stackrel{\textstyle
g_{r+s}}{\longrightarrow} \cdot\ b_m^\dag, \label{eq:comm28}\\
{}\Bigl[a_j^\dag\ \cdot\stackrel{\textstyle
\check{G}_r}{\longrightarrow}\cdot\  a_k,\,
a_\ell^\dag\ \cdot\stackrel{\textstyle
\check{G}_s}{\longrightarrow}\cdot\  a_m\Bigr] & = &
\delta_{k\ell}\,a_j^\dag\ \cdot\stackrel{\textstyle \check{G}_{r+s}}
{\longrightarrow} \cdot\ a_m - \delta_{jm}\,a_\ell^\dag\ \cdot
\stackrel{\textstyle \check{G}_{r+s}}{\longrightarrow} \cdot\ a_k,
\label{eq:comm29}\\
{}\Bigl[b_j^\dag\ \cdot\stackrel{\textstyle \widehat{G}_r}{\longrightarrow}
\cdot\ b_k,\, a_\ell^\dag\ \cdot\stackrel{\textstyle
g_s}{\rightarrow}\cdot\ b_m^\dag\Bigr] & = &
\delta_{km}\,a_\ell^\dag\ \cdot\stackrel{\textstyle
g_{r+s}}{\longrightarrow} \cdot\ b_j^\dag, \label{eq:comm30}\\
{}\Bigl[b_j^\dag\ \cdot\stackrel{\textstyle
\widehat{G}_r}{\longrightarrow}\cdot\ b_k,\, b_\ell^\dag\
\cdot\stackrel{\textstyle \widehat{G}_s}{\longrightarrow}\cdot\ b_m\Bigr] & =
& \delta_{k\ell}\,b_j^\dag\ \cdot\stackrel{\textstyle \widehat{G}_{r+s}}
{\longrightarrow} \cdot\ b_m - \delta_{jm}\,b_\ell^\dag\ \cdot
\stackrel{\textstyle \widehat{G}_{r+s}}{\longrightarrow} \cdot\ b_k.
\label{eq:comm31}
\end{eqnarray}

\subsection{Commutators of powers of operators}
To evaluate powers of operators, we shall need some repeated commutators.

\vspace{3pt}

\noindent\textbf{Lemma}: For any positive integers $r,s,q$,
\begin{equation}
{}\Bigl[\Bigl[a_j\ \cdot\stackrel{\textstyle g^*_r}{\longrightarrow}\cdot\
b_k,\, a_\ell^\dag\ \cdot\stackrel{\textstyle g_s}{\longrightarrow}\cdot\ 
b_m^\dag\Bigr],\, a_\ell^\dag\ \cdot\stackrel{\textstyle
g_q}{\longrightarrow}\cdot\ b_m^\dag\Bigr]
=2\delta_{j\ell}\delta_{km}\,a_\ell^\dag\ \cdot\stackrel{\textstyle
g_{q+r+s-1}} {\longrightarrow}\cdot\ b_m^\dag.
\end{equation}
It is worth noticing that this double commutator is of \textit{creation
type}, and thus commutes with all other operators of that type.
The commutation relations also imply 

\vspace{3pt}

\noindent\textbf{Lemma}: Regardless of
what functions decorate the arrows,
\begin{equation}
{}[[a\ \cdot\rightarrow \cdot\ b_j,\, a^\dag\
\cdot\rightarrow\cdot\ 
  b_k^\dag],\, a^\dag\ \cdot\rightarrow\cdot\ b_k^\dag] =
0,\quad
\text{if\ }j \ne k.
\label{eq:ABB}
\end{equation}
In particular the operators
$a_j\ \cdot\stackrel{\textstyle g^*}{\rightarrow}\cdot\ b_k$ and
$a_j^\dag\ \cdot\stackrel{\textstyle g}{\rightarrow}\cdot\ b^\dag_k$ above
fulfill the conditions for this lemma.

\vspace{3pt}

\noindent\textbf{Proposition}: Write $(ab_j)$ for
$a\ \cdot\stackrel{\textstyle g^*}{\rightarrow}\cdot\ b_j$ and for
non-negative integers $n_1,n_2,\ldots,n_K$, let $n = \sum_{j=1}^K
n_j$; then
\begin{equation} 
\langle 0|\left(\prod_{j=1}^K (ab_j)^{n_j}\right)\left(\prod_{j=1}^K
  (ab_j)^{\dag n_j}\right)|0\rangle =
  \frac{n_1!n_2!\cdots n_K!}{n!}\,\langle 0|(ab_1)^n(ab_1)^{\dag
n}|0\rangle.
\label{eq:nktangle}
\end{equation}

\noindent\textit{Proof}: The proof employs the symmetry operator of
Eq. (\ref{eq:Sdef}). The left side of the equation is the inner
product of a vector $|\phi\rangle$ with itself, where
\begin{eqnarray}
|\phi\rangle & = & h_{2n}\!:\!\mbox{Pol}^\dag |0\rangle \nonumber \\ &
= & \int d\omega_1\,d\tilde{\omega}_1\cdots d\omega_n\,
d\tilde{\omega}_n\,h_{2n}(\omega_1,\tilde{\omega}_1,\ldots,\omega_n,
\tilde{\omega}_n) \mbox{Pol}^\dag |0\rangle ,\nonumber
\end{eqnarray}
with
\begin{equation}
h_{2n}(\omega_1,\tilde{\omega}_1,\ldots,\omega_n,\tilde{\omega}_n)
= g(\omega_1,\tilde{\omega}_1)\cdots g(\omega_n,\tilde{\omega}_n)
\label{eq:h2nDef}
\end{equation}
and 
\begin{equation}
\mbox{Pol}^\dag =
\left(\prod_{j=1}^{n_1}a^\dag(\omega_j)b^\dag_1(\tilde{\omega}_j)\right)
\cdots
\left(\prod_{j=n-n_K+1}^{n}a^\dag(\omega_j)b^\dag_K(\tilde{\omega}_j)
\right).
\end{equation}
Because of the commutativity relations, the integral is unchanged 
under certain permutations of the arguments of $h_{2n}$.
Interchanging the operators $a^\dag(\omega_j)$ among themselves
yields
\begin{equation} 
|\phi\rangle =
 (\mathcal{S}(\omega_1,\ldots,\omega_n)h_{2n})\!:\!\mbox{Pol}^\dag
 |0\rangle.
\end{equation}
In addition, letting $\vec{\omega}_j = (\omega_j,\tilde{\omega}_j)$,
we have 
\begin{equation}
|\phi\rangle = (\mathcal{S}(\vec{\omega}_1,\ldots,\vec{\omega}_n)
 h_{2n})\!:\!\mbox{Pol}^\dag |0\rangle.\label{eq:C39}
\end{equation}
These two symmetries for $h_{2n}$ and its product form defined in Eq.\
(\ref{eq:h2nDef}) imply a third symmetry:
\begin{equation}
|\phi\rangle =
 (\mathcal{S}(\tilde{\omega}_1,\ldots,\tilde{\omega}_n)h_{2n})
\!:\!\mbox{Pol}^\dag|0\rangle.
\end{equation}
{}From this symmetry and Lemma (\ref{eq:commnn}) the proposition follows.
$\Box$

\noindent This generalizes Lemma (\ref{eq:nkrelation}).

We shall need to refer to a function
\begin{equation}
\Xi_g(n) \defeq  \frac{1}{n!}\,\langle 0|(ab)^n(ab)^{\dag
n}|0\rangle.
\label{eq:XiDefcopy}
\end{equation}
This function is independent of the choice of operators $a(\omega)$
and $b(\omega)$ as long as each operator satisfies the commutation
relations Eqs.\ (\ref{eq:comm0}), (\ref{eq:comm1}) and the two operators
are mutually orthogonal.
\vspace{3pt}

\noindent\textbf{Proposition}: Write $(ab)$ for
$a\ \cdot\stackrel{\textstyle g^*}{\rightarrow}\cdot\ b$.  Then
\begin{eqnarray} 
\Xi_g(n) & \defeq  &
\frac{1}{n!}\,\langle 0|(ab)^n(ab)^{\dag n}|0\rangle \nonumber \\ & = &
\sum^n_{{\scriptstyle \nu_1,\ldots,\nu_n=0}\atop{\scriptstyle \sum
k\nu_k = n}}
\frac{n!}{1^{\nu_1}\nu_1!2^{\nu_2}\nu_2!\cdots
n^{\nu_n}\nu_n!}\prod_{j=1}^n[\mbox{Loop}(\widehat{G}_j)]^{\nu_j},
\label{AnAnGen}
\end{eqnarray} where the $\nu_k$ are restricted as indicated.

\vspace{5pt}

\noindent\textit{Proof}: From symmetry considerations
and Lemma (\ref{eq:commnn}) we get
\begin{eqnarray} \frac{1}{n!}\,\langle 0|(ab)^n(ab)^{\dag n}|0\rangle &
= & \frac{1}{n!}\int d\omega_1\,d\tilde{\omega}_1\cdots d\omega_n\,
d\tilde{\omega}_n\, d\omega'_1\,d\tilde{\omega}'_1\cdots d\omega'_n
d\tilde{\omega}'_n \nonumber \\ & &[g(\omega_1,\tilde{\omega}_1)\cdots
g(\omega_n,\tilde{\omega}_n)]\,[g^*(\omega'_1,\tilde{\omega}'_1)\cdots
g^*(\omega'_n,\tilde{\omega}'_n)] \nonumber \\ & &{}\times
\left(\sum_{\pi
\in S_n}\prod_{j=1}^n\delta(\omega_j - \omega'_{\pi j})
\right)\left(\sum_{\pi \in S_n}\prod_{j=1}^n\delta(\tilde{\omega}_j -
\tilde{\omega}'_{\pi j}) \right).
\end{eqnarray}  
By virtue of Eq.\ (\ref{eq:C39}), any permutation of the variables
$\tilde{\omega}'_{\pi j}$ is compensated out by a corresponding
permutation of the variables $\omega'_{\pi j}$, which implies
\begin{eqnarray} 
\frac{1}{n!}\,\langle 0|(ab)^n(ab)^{\dag
n}|0\rangle
 & = &  \int d\omega_1\,d\tilde{\omega}_1\cdots
d\omega_n\, d\tilde{\omega}_n\, d\omega'_1\cdots d\omega'_n
\,[g(\omega_1,\tilde{\omega}_1)\cdots
g(\omega_n,\tilde{\omega}_n)]\nonumber \\ &
&\times [g^*(\omega'_1,\tilde{\omega}_1)\cdots
g^*(\omega'_n,\tilde{\omega}_n)] 
\left(\sum_{\pi
\in S_n}\prod_{j=1}^n\delta(\omega_j - \omega'_{\pi j})
\right)\nonumber\\
&=& n!\int d\omega_1d\tilde{\omega}_1\cdots d\omega_nd\tilde{\omega}_n\,
|\mathcal{S}(\omega_1,\ldots,\omega_n) g(\omega_1,\tilde{\omega}_1)\cdots
g(\omega_n,\tilde{\omega}_n)|^2.\nonumber\\
\end{eqnarray}
The effect of each of these remaining permutations is to generate a
product of integrals, each of convolutions of $g$'s with matching
convolutions of $g^*$'s, according to the cycle structure of the
permutation group $S_n$.  The convolutions generated by a permutation
are those that correspond to the cycles of its conjugacy class. Each
conjugacy class is characterized by some $[\nu_1,\ldots,\nu_n]$ where
$\nu_1$ is the number of one-cycles, $\nu_2$ is the number of
two-cycles, etc.\ (with $\sum_{k=1}^n k \nu_k = n$)
\cite{sternberg}. The number of permutations in a conjugacy
class is just that stated in the Proposition
\cite{sternberg}. $\Box$

Examples are 
\begin{eqnarray}
\Xi_g(0) & = & 1, \nonumber \\
\Xi_g(1) & = & \mbox{Loop}(\widehat{G}_1), \nonumber \\ \Xi_g(2) & = &
[\mbox{Loop}(\widehat{G}_1)]^2 + \mbox{Loop}(\widehat{G}_2),
\nonumber \\ \Xi_g(3) & = & [\mbox{Loop}(\widehat{G}_1)]^3 +
3[\mbox{Loop}(\widehat{G}_1)][\mbox{Loop}(\widehat{G}_2)] +
2\mbox{Loop}(\widehat{G}_3). 
\end{eqnarray}
As a check on the Proposition, these
examples can also be demonstrated by repeated use of Eq.\
(\ref{eq:AnAn}) along with the commutation relations Eqs.\
(\ref{eq:comm27})--(\ref{eq:comm31}).

We will need to deal with partial traces, as defined in Sec.~\ref{sec:2}, 
in particular we need
\begin{eqnarray}\lefteqn{
\mbox{Tr}_a[(ab)^{\dag n}|0\rangle\langle 0|(ab)^n] =
   \int\!\!d\tilde{\bm{\omega}}d\tilde{\bm{\omega}}' } \nonumber \\
   &&\left[\int\!\!d\bm{\omega}d\bm{\omega}'
   \bm{g}(\bm{\omega},\tilde{\bm{\omega}})
   \bm{g}^*(\bm{\omega}',\tilde{\bm{\omega}}') \langle
   0_a|a^n(\bm{\omega}')a^{\dag n}(\bm{\omega})|0_a\rangle\right]
   b^{\dag n}(\tilde{\bm{\omega}})|0_b\rangle\langle
   0_b|b^n(\tilde{\bm{\omega}}') \nonumber \\ & = &
   n!\int\!\!d\tilde{\bm{\omega}}d\tilde{\bm{\omega}}'
   \left[\int\!\!d\bm{\omega}\,\bm{g}(\bm{\omega},\tilde{\bm{\omega}})
   \mathcal{S}(\bm{\omega})\bm{g}^*(\bm{\omega},\tilde{\bm{\omega}}')
   \right] b^{\dag n}(\tilde{\bm{\omega}})|0_b\rangle\langle
   0_b|b^n(\tilde{\bm{\omega}}'),
\end{eqnarray}
where we have used the compact
   notation defined in Sec.\ \ref{sec:2}, with
   $\bm{g}(\bm{\omega},\tilde{\bm{\omega}})\stackrel{\rm def}{=}
   g(\omega_1,\tilde{\omega}_1)\cdots$ $g(\omega_n,\tilde{\omega}_n)$,
and the second equality invokes Lemma (B15).  Because of symmetry
of the $b$ operators under permutations of their arguments, this
simplifies to 
\begin{eqnarray}
\mbox{Tr}_a[(ab)^{\dag n}|0\rangle\langle 0|(ab)^n] =
   n!\int\!\!d\tilde{\bm{\omega}}d\tilde{\bm{\omega}}' 
\left[\int\!\!d\bm{\omega}\,
\bm{g}(\bm{\omega},\tilde{\bm{\omega}})
   \bm{g}^*(\bm{\omega},\tilde{\bm{\omega}}') \right]
   b^{\dag n}(\tilde{\bm{\omega}})|0_b\rangle\langle
   0_b|b^n(\tilde{\bm{\omega}}'). 
\label{eq:pTrace0} \nonumber \\ &&
\end{eqnarray}
Writing out the compact notion, we note that the inner integral above
is
\begin{eqnarray}
\int\!\!d\bm{\omega}\,
\bm{g}(\bm{\omega},\tilde{\bm{\omega}})
\bm{g}^*(\bm{\omega},\tilde{\bm{\omega}}')& = &
\prod_{j=1}^n\left(\int\!\!d\omega\,
g(\omega,\tilde{\omega}_j)g^*(\omega,\tilde{\omega}'_j)\right)
\nonumber \\ &=& \prod_{j=1}^n\left(\tilde{\omega}_j
\stackrel{{\textstyle g}}{\leftarrow}\cdot\stackrel{{\textstyle g^*}}\rightarrow
\tilde{\omega}'_j \right).
\end{eqnarray}
As a check, we note that the trace over the $b$-mode of Eq.\
(\ref{eq:pTrace0}) results in Proposition (C42), as it should.

\subsection{Examples of frequency functions}
We consider a family of functions $g_\zeta(\omega,\tilde{\omega})$ and
show two limiting cases. For any real-valued functions $\phi(\omega)$
and $\tilde{\phi}(\tilde{\omega})$ and positive real parameters  
$\sigma$ and $\tilde{\sigma}$, let
\begin{equation}
g_\zeta(\omega,\tilde{\omega}) =
\frac{1}{\sqrt{\sigma\tilde{\sigma}}}\,e^{i\phi(\omega)}
e^{i\tilde{\phi}(\tilde{\omega})} F\!\left(\zeta;\frac{\omega-
\omega_0}{\sigma},\frac{\tilde{\omega}-\tilde{\omega}_0}
{\tilde{\sigma}}\right),
\label{eq:gzet}
\end{equation}
where we define
\begin{eqnarray}
{}F(\zeta;x,y) & \defeq  &
\sqrt{\frac{2}{\pi}}\exp\left\{-\frac{1}{2}\left[\left(
\sqrt{\zeta^2+1}+\zeta\right)(x+y)^2 +
\left(\sqrt{\zeta^2+1}+\zeta\right)^{-1}(x-y)^2\right]\right\} \nonumber
\\  & = &
\sqrt{\frac{2}{\pi}}\exp\Bigl\{-\Bigl[\sqrt{\zeta^2+1}\,(x^2 + y^2) + 2
\zeta x y\Bigr]\Bigr\};
\label{eq:FDefapp}
\end{eqnarray}
regardless of the value of $\zeta$,
\begin{equation}
\int^\infty_{-\infty} dx\,dy\, |F(\zeta;x,y)|^2 = 1.
\end{equation}
Thus for any choice of center frequencies $\omega_0$ and
$\tilde{\omega}_0$, bandwidth parameters $\sigma$ and
$\tilde{\sigma}$, and phase functions $\phi(\omega)$ and
$\tilde{\phi}(\tilde{\omega})$, we get a family of $g_\zeta$'s.

By changing integration variables, one shows
\begin{equation}
\mbox{Loop}\biggl[\biggl(\stackrel{\textstyle g_\zeta}{\rightarrow} \cdot
 \stackrel{\textstyle g^*_\zeta}{\leftarrow}\biggr)^n\biggr] =
 \mbox{Loop}\biggl[\biggl(\stackrel{\textstyle F}{\rightarrow} \cdot
 \stackrel{\textstyle F^*}{\leftarrow}\biggr)^n\biggr],
\end{equation}
and similarly  for the convolution of $g_\zeta$'s
\begin{equation} \omega
 \biggl(\stackrel{\textstyle g_\zeta}{\rightarrow} \cdot
 \stackrel{\textstyle g^*_\zeta}{\leftarrow}\biggr)^n \cdot
 \stackrel{\textstyle g_\zeta}{\rightarrow} \tilde{\omega} =
 \frac{1}{\sqrt{\sigma\tilde{\sigma}}}\,e^{i\phi(\omega)}
e^{i\tilde{\phi}(\tilde{\omega})}
 \left(\frac{\omega -
 \omega_0}{\sigma}\biggl(\stackrel{\textstyle F}{\rightarrow} \cdot
 \stackrel{\textstyle F^*}{\leftarrow}\biggr)^n \cdot
 \stackrel{\textstyle F}{\rightarrow}\frac{\tilde{\omega}-\tilde{\omega}_0}
 {\tilde{\sigma}}\right).
\label{eq:gconvo}
\end{equation}
It remains to compute the convolution integrals for
$F$. Although $g_\zeta(\omega,\tilde{\omega})$ need be neither
real-valued nor symmetric in $\omega$ and $\tilde{\omega}$, the
function $F(\zeta;x,y)$ is real and symmetric, implying
\begin{equation}
\stackrel{\textstyle F}{\rightarrow}\; =\; \stackrel{\textstyle F}{\leftarrow}\;
 = \;\stackrel{\textstyle F^*}{\rightarrow}\;=
 \;\stackrel{\textstyle F^*}{\leftarrow}.
\label{eq:fsym}
\end{equation}
{}For this reason arrow expressions built up from convolutions of
factors of $F$ are invariant under any number of reverses of
arrow directions; to emphasize this indifference to arrow direction in
$F$ (but not $g_\zeta$), we write
\begin{equation}
\stackrel{\textstyle F}{\longleftrightarrow}.
\end{equation}
To compute the convolution integrals for $F$, consider a sequence of
$\zeta_j$.  Abbreviating $F(\zeta_j;\mathbf\cdot,\mathbf\cdot)$ by
$F_j$ and $\sqrt{\zeta_j^2+1}$ by $c_j$, we have
\begin{eqnarray}\lefteqn{
x \stackrel{\textstyle F_j}{\longleftrightarrow}\cdot
\stackrel{\textstyle F_k}{\longleftrightarrow} y  }\quad
\nonumber \\ &\defeq  & \frac{2}{\pi}\int
dz\,\exp\{-[c_j(x^2+z^2)+c_k(z^2+y^2) + 2\zeta_jxz + 2\zeta_kzy]\}
\nonumber \\ & = & \sqrt{\frac{2}{c_j+c_k}}\,
\sqrt{\frac{2}{\pi}}\exp\left\{-\left(\Biggl[\left(-
\frac{\zeta_j\zeta_k}{c_j+c_k}\right)^2 + 1\Biggr]^{1/2} (x^2+y^2)-
\frac{2\zeta_j\zeta_k}{c_j+c_k}\,xy\right)\right\} \nonumber \\ & = &
\sqrt{\frac{2}{c_j+c_k}} \,F\!\left(-
\frac{\zeta_j\zeta_k}{c_j+c_k};x,y\right).
\label{eq:convo}
\end{eqnarray}
Setting $y=x$ and integrating yield
\begin{equation}
(\forall\ j)\quad \mbox{Loop}\biggl(\stackrel{\textstyle
F_j}{\longleftrightarrow}\cdot
\stackrel{\textstyle F_j}{\longleftrightarrow}\biggr) = 1.
\label{eq:loop1}
\end{equation}
{}From this and Eqs.\ (\ref{eq:FDefapp}) and (\ref{eq:convo})
follows the
\vspace{3pt}

\noindent\textbf{Lemma}: 
\begin{equation}
x \biggl(\stackrel{\textstyle F(\zeta)}{\longleftrightarrow}\biggr)^ny =
\sqrt{\kappa(\zeta,n)}\,F(\zeta_n;x,y),
\label{eq:knlemma}
\end{equation}
where we have written $F(\zeta)$ as an abbreviation for
$F(\zeta;\cdot,\cdot)$ and 
\begin{eqnarray}
\kappa(\zeta,1) & = & 1,\label{eq:kapneqone} \\ \zeta_1 & = & \zeta, \\
\zeta_{n+1} & = & -
\frac{\zeta_1\zeta_n}{\sqrt{\zeta_1^2+1}+\sqrt{\zeta_n^2+1}}, \\
\kappa(\zeta,n+1) & = &
\frac{2}{\sqrt{\zeta_1^2+1}+\sqrt{\zeta_n^2+1}}.
\label{eq:kformula}
\end{eqnarray} 
{}From the lemma and these equations we find
\begin{eqnarray}
\mbox{Loop}\Biggl(\stackrel{\textstyle \check{G}_{\zeta
n}}{\longrightarrow}\Biggr) & \equiv &
\mbox{Loop}\Biggl[\Biggl(\stackrel{\textstyle g_\zeta}{\rightarrow}
\cdot\stackrel{\textstyle g^*_\zeta} {\leftarrow}\Biggr)^n\Biggr]
\nonumber
\\ & = &
\mbox{Loop}\Biggl[\Biggl(\stackrel{\textstyle F(\zeta)}{\longleftrightarrow}
\cdot\stackrel{\textstyle F(\zeta)} {\longleftrightarrow}\Biggr)^n\Biggr]
\nonumber \\ & = &
\mbox{Loop}\Biggl[\Biggl(\stackrel{\textstyle
F(\zeta)}{\longleftrightarrow}\Biggr)^n\cdot \Biggl(\stackrel{\textstyle
F(\zeta)}{\longleftrightarrow}\Biggr)^n\Biggr] \nonumber \\ & = &
\kappa(\zeta,n),
\end{eqnarray}
with $\kappa(\zeta,n)$ as defined in (\ref{eq:kformula}).  Similarly this
and Eq.\ (\ref{eq:gn2Def}) yield
\begin{equation}
\text{Loop}\biggl(\stackrel{\textstyle g_{\zeta n}}{\longrightarrow}
\cdot\stackrel{\textstyle g^*_{\zeta n}} {\longleftarrow}\biggr) =
\text{Loop}\Biggl(\stackrel{\textstyle
\check{G}_{\zeta,2n-1}}{\longrightarrow}\Biggr) = \kappa(\zeta,2n-1).
\label{eq:gznNorm}
\end{equation}
Note that Eq.\ (\ref{eq:kformula}) implies 
\begin{equation}
(\forall\ n \ge 2)\quad \kappa(\zeta,n) \le \frac{2}{|\zeta|}.
\label{eq:ksqbound}
\end{equation}
For more efficient calculation, define $R(\zeta,n)$ for $n \ge 2$ by
\begin{equation}
\kappa(\zeta,n) = \frac{R(\zeta,n)}{\sqrt{\zeta^2+1}}.
\label{eq:RzetDef}
\end{equation} 
{}From Eqs.\ (\ref{eq:kapneqone}) and (\ref{eq:kformula}), it follows
that $R(\zeta,n)$ is a rational function of $\zeta$ satisfying the
following recursion relation

\vspace{8pt}
\noindent\textbf{Lemma}:\vadjust{\kern-8pt} 
\begin{eqnarray} 
R(\zeta,2) & = & 1, \\
(\forall\ n \ge 2)\quad R(\zeta,n+1) & = &  
\left( 1-\frac{\zeta^2}{4(\zeta^2+1)}\,R(\zeta,n) \right)^{-1}.
\end{eqnarray}
{}From Lemma (\ref{eq:knlemma}) and Eq.\ (\ref{eq:gconvo}) one also finds
\begin{eqnarray}
\omega\stackrel{\textstyle g_{\zeta n}}{\longrightarrow}\tilde{\omega} & = &
 k_{2n+1}\,\frac{1}{\sqrt{\sigma\tilde{\sigma}}}\,e^{i\phi(\omega)}
 e^{i\tilde{\phi}(\tilde{\omega})} \,
 F\!\left(\zeta_{2n+1};\frac{\omega -
 \omega_0}{\sigma},\frac{\tilde{\omega}-\tilde{\omega}_0}
 {\tilde{\sigma}}\right),
\label{eq:gzetN} \\
\omega_1\stackrel{\textstyle \check{G}_{\zeta n}}{\longrightarrow}\omega_2 & = &
\sigma^{-1}k_{2n} \exp\{i[\phi(\omega_1)-\phi(\omega_2)]\}
\,F\!\left(\zeta_{2n};\frac{\omega_1 -
\omega_0}{\sigma},\frac{\omega_2-\omega_0} {\sigma}\right),
\label{eq:Gcheck} \\ \tilde{\omega}_1\stackrel{\textstyle
\widehat{G}_{\zeta n}}{\longrightarrow}\tilde{\omega}_2 & = & 
\tilde{\sigma}^{-1} k_{2n}
\exp\{i[\tilde{\phi}(\tilde{\omega}_1) - \tilde{\phi}(\tilde{\omega}_2)]\}
\,F\!\left(\zeta_{2n};\frac{\tilde{\omega}_1 -
\tilde{\omega}_0}{\tilde{\sigma}},\frac{\tilde{\omega}_2-\tilde{\omega}_0}
{\tilde{\sigma}}\right).
\end{eqnarray}

\subsection{Limiting cases}
\noindent\textbf{Case I}.\ \ No frequency entanglement:
$g_{\mathrm I}(\omega,\tilde{\omega})$ $= f(\omega)h(\tilde{\omega})$,
normalized so that $\int\!\!\int d\omega\,
d\tilde{\omega}\,|g(\omega,\tilde{\omega})|^2 =
\int d\omega\,|f(\omega)|^2 =
\int d\tilde{\omega}\,|h(\tilde{\omega})|^2 = 1$.  In this case, one
skips the fancy commutation relations because the operators
all factor; we find

\vspace{3pt}

\noindent\textbf{Lemma}:
\begin{equation}
\Xi_{\mathrm I}(n) \defeq \frac{1}{n!}\,\langle
0|\Bigl(a\ \cdot\stackrel{\textstyle g^*_{\mathrm
I}}{\longrightarrow}\cdot\ b\Bigr)^n \Bigl(a^\dag\
\cdot\stackrel{\textstyle g_{\mathrm I}}{\longrightarrow}\cdot\
b^\dag\Bigr)^n |0\rangle = n!,
\label{eq:XiI}
\end{equation}
which follows from the discussion of broad-band coherent states in
Sec.~\ref{subsec:2C}.

\noindent\textbf{Case II}.\ \ Limit as $\zeta \rightarrow \pm \infty$.  It
makes no sense to ask for the limit of $g_\zeta$; however, we explore
large values of $|\zeta|$ by looking at the limit of the commutation
relations.  From Eqs.\ (\ref{eq:kformula}) and (\ref{eq:gznNorm}) we see
for this limit
\begin{equation} 
(\mbox{for }n \ge 2)\quad \lim_{\zeta \rightarrow \pm \infty} k_n =
\lim_{\zeta \rightarrow \pm \infty}
\mbox{Loop}\biggl(\stackrel{\textstyle
\check{G}_{\text{II}n}}{\longrightarrow}\biggr) =
\lim_{\zeta \rightarrow \pm \infty}
\mbox{Loop}\biggl(\stackrel{\textstyle g_{\zeta n}}{\longrightarrow}
\cdot\stackrel{\textstyle g^*_{\zeta n}} {\longleftarrow}\biggr) = 0,
\end{equation} resulting in specializing the commutation
Eqs.\ (\ref{eq:comm27})--(\ref{eq:comm31}), for sufficiently
large $|\zeta|$, to
\begin{eqnarray}
\lefteqn{\Bigl[a_j\ \cdot\stackrel{\textstyle g^*}{\rightarrow}\cdot\ b_k,\,
a_\ell^\dag\ \cdot\stackrel{\textstyle g}{\rightarrow}\cdot\
b_m^\dag\Bigr]}\quad\nonumber\\ 
& = & \delta_{km}\, a_\ell^\dag\ \cdot\stackrel{\textstyle
\check{G}_1}{\longrightarrow}
\cdot\ a_j +\delta_{j\ell}\, b_m^\dag\ \cdot
\stackrel{\textstyle \widehat{G}_{1}}{\longrightarrow} \cdot\ b_k +
\delta_{j\ell}\delta_{km},
\label{eq:commII27}
\end{eqnarray}
\begin{eqnarray}
\lim_{\zeta \rightarrow \pm \infty}
\biggl[a_j^\dag\ \cdot\stackrel{\textstyle
\check{G}_1}{\longrightarrow}\cdot\ a_k,
a_\ell^\dag\ \cdot\stackrel{\textstyle g}{\rightarrow}\cdot\
b_m^\dag\biggr] & = & 0,
\label{eq:commII28}\\ \lim_{\zeta \rightarrow \pm \infty}
\biggl[a_j^\dag\ \cdot\stackrel{\textstyle
\check{G}_1}{\longrightarrow}\cdot\ a_k,
a_\ell^\dag\ \cdot\stackrel{\textstyle
\check{G}_1}{\longrightarrow}\cdot\ a_m\biggr] & = & 0,
\label{eq:commII29}\\
\lim_{\zeta \rightarrow \pm
\infty}\biggl[b_j^\dag\ \cdot\stackrel{\textstyle
\widehat{G}_1}{\longrightarrow}\cdot\ b_k,
a_\ell^\dag\ \cdot\stackrel{\textstyle g}{\rightarrow}\cdot\
b_m^\dag\biggr] & = & 0,
\label{eq:commII30}\\
\lim_{\zeta \rightarrow \pm \infty}
\biggl[b_j^\dag\ \cdot\stackrel{\textstyle
\widehat{G}_1}{\longrightarrow}\cdot\ b_k,
b_\ell^\dag\ \cdot\stackrel{\textstyle \widehat{G}_1}{\longrightarrow}\cdot\ 
b_m\biggr] & = &0.
\label{eq:commII31}
\end{eqnarray}
In this case, the double commutator
$\Bigl[\Bigl[a_j\ \cdot\stackrel{\textstyle g^*}{\rightarrow}\cdot\ b_k,
\,a_\ell^\dag\ \cdot\stackrel{\textstyle g}{\rightarrow}\cdot\
b_m^\dag\Bigr],\, a_\ell^\dag\ \cdot\stackrel{\textstyle
g}{\rightarrow}\cdot\ b_m^\dag\Bigr]$ is effectively zero, so that Lemma
(\ref{eq:lem33}) applies.

{}From Lemma (\ref{eq:lemComII}) and Eqs.\
(\ref{eq:commII27})--(\ref{eq:commII31}) follows the corresponding rule
for evaluating Case-II operator products:

\vspace{3pt}

\noindent\textbf{Lemma}:
\begin{equation}
\Xi_{\mathrm{II}}(n) \defeq 
\lim_{\zeta \rightarrow \pm \infty}
\frac{1}{n!}\,\langle 0|\Bigl(a\ \cdot\stackrel{\textstyle
g^*}{\rightarrow}\cdot\ b\Bigr)^n \Bigl(a^\dag\ \cdot\stackrel{\textstyle
g}{\rightarrow}\cdot\ b^\dag\Bigr)^n |0\rangle = 1. 
\label{eq:XiII}
\end{equation}

\section{Fourier transforms in space and time}\label{app:D}
Let $f(x,t)$ be any operator-valued function for which Fourier
transforms make sense, and define the Fourier transform pair:
\begin{eqnarray}
\overline{f}(\omega,k) & = & (2\pi)^{-1}\int_{-\infty}^{\infty}dt\, 
\int_{-\infty}^{\infty} dx\, f(x,t)e^{i(\omega t +
  k x)}, \\ f(x,t) & = & (2\pi)^{-1}\int_{-\infty}^{\infty}d\omega\,
\int_{-\infty}^{\infty} dk\, \overline{f}(\omega,k)e^{-i(\omega t +
  k x)}. \label{eq:inv}
\end{eqnarray}
On taking the adjoint of these equations, one sees that:

\vspace{3pt}

\noindent\textbf{Lemma}: The Fourier transform of
$f^\dag(x,t)$ is related to the adjoint of the transform of $f(x,t)$
by
\begin{equation}
\overline{f}^\dag(-\omega,-k)= (\overline{f}(\omega,k))^\dag.
\label{eq:lem9a}\end{equation}

\vspace{3pt}

\noindent\textbf{Lemma}:  If $f(x,t) = f^\dag(x,t)$, then
\begin{equation}
\overline{f}(-\omega,-k)= (\overline{f}(\omega,k))^\dag,
\end{equation} so that the Fourier transform of a hermitian
operator function is specified for the whole ($\omega,k$)-plane once
it is specified for any half-plane touching the origin.  In particular
Eq.\ (\ref{eq:inv}) can be replaced by
\begin{equation}
f(x,t)  =  (2\pi)^{-1}\int_{-\infty}^{\infty}d\omega\,
\int_0^{\infty} dk\, \left(\overline{f}(\omega,k)e^{-i(\omega t +
  k x)}+ \overline{f}^\dag(\omega,k)e^{i(\omega t +
  k x)}\right),
\end{equation}
where the integrand is integrated over the half-plane $k > 0$;
alternatively $f(x,t)$ can be expressed by the same integrand
integrated over the half-plane defined by $\omega > 0$.  More
generally, the region of integration needs to be obtained from the
region excluded by reflection through the origin; thus the region of
integration need not be aligned with the axes and need not have a
straight boundary.
\vspace{3pt}

\noindent\textbf{Lemma}: If $\overline{f}(\omega,k) =
\overline{f}^\dag(\omega,k)$, then
\begin{equation}
\overline{f}(-\omega,-k)= (\overline{f}(\omega,k))^\dag.
\end{equation}

Now consider the form of a Fourier transform of any solution to the wave
equation
\begin{equation}(\partial_x^2 - \partial_t^2)g(x,t) = 0;
\label{eq:wave}
\end{equation}
this equation has as its general solution 
\begin{equation}g(x,t) = g_+(t-x)+g_+(t+x),
\end{equation} so that the transform of $g$ has the form
\begin{equation}
\overline{g}(\omega,k) =
\sqrt{2\pi}\,[\delta(k+\omega)\overline{g}_+(\omega) +
  \delta(k-\omega)\overline{g}_-(\omega)],
\label{eq:wvxf}
\end{equation}
where
\begin{equation}\overline{g}_\pm(\omega)\defeq 
\frac{1}{\sqrt{2\pi}}\int_{-\infty}^\infty du\, e^{i\omega u}g_\pm(u).
\end{equation}
Taking the inverse transform of Eq.\ (\ref{eq:wvxf}), one obtains for
the general solution to Eq.\ (\ref{eq:wave})
\begin{equation}g(x,t) = \frac{1}{\sqrt{2\pi}}\int_{-\infty}^\infty
d\omega\,[\overline{g}_+(\omega)e^{-i\omega(t-x)} +
\overline{g}_-(\omega)e^{-i\omega(t+x)}].
\label{eq:wvxf2}
\end{equation}

\section{Expansion of light states in tensor products of broad-band
coherent states}\label{app:E}

Consider the subspace of broad-band coherent states spanned by
$a_f^{\dag n}|0\rangle$, $n = 0, 1, 2$, \dots. It follows from
Louisell [\onlinecite{louisell}, p.~106] that the unit operator on this
subspace is
\begin{equation}
\int |\alpha,a_f\rangle \langle \alpha,a_f|\,\frac{d^2 \alpha}{\pi}.
\end{equation}
Next, let $F$ be any set of orthonormal functions
$f_j(\omega)$, $j = 1,2$, \dots.  This implies that the set of
operators $|\alpha_j, a_{f_j}\rangle\langle \alpha_j, a_{f_j}|$, $j =
1,2$, \dots, are mutually orthogonal projections.  We say a set of
light states is `coherently expressible with respect to $F$' if for
some set $F$ of orthonormal functions $f_j(\omega)$, $j = 1,2$, \dots,
every state of the set is some weighted sum (or integral) over $j$ and
$\alpha_j$ of states of the form
\begin{equation}
\prod_j |\alpha_j,a^\dag_{f_j}\rangle,
\end{equation}
where the product is a tensor product.  The unit operator for the
vector space of such states coherently expressible with respect to $F$
is then
\begin{equation}
\sum_j \int |\alpha_j,a_{f_j}\rangle \langle \alpha,a_{f_j}|\,\frac{d^2
\alpha_j}{\pi}.
\end{equation}
Subtleties of coherent states in infinite dimensional vector spaces
are touched on in [\onlinecite{jauch}, pp.\ 503--512].

\section{MATLAB programs for Section 10}\label{app:F} 

Here we record the MATLAB scripts and functions used to generate Fig.\
\ref{fig:7}.

\vspace{5pt}\noindent (1) \texttt{\,Partfn.m\,} starts the calculation by
preparing a list
\texttt{\,yList\,} of the partitions of integers needed for Eq.\
(\ref{AnAnGen}).  It stores \texttt{\,yList\,} in a file
\texttt{\,Part.mat\,}.  It needs to be run only once, with a value of
\texttt{\,nmax = floor(3 * mu\_max + 16)\,}, where \texttt{\,mu\_max\,} is
the highest value of $\mu$ covered. (For \texttt{\,nmax = 32\,},
\texttt{\,Partfn.m\,} takes 20 minutes on a Pentium-4 desktop computer.)

\vspace{5pt}\noindent (2) \texttt{\,muRun.m\,} generates data in the cell variable
\texttt{\,muResult\,} for later plotting.  It has a section that is
easily modified in order to zoom in on one or another parameter
region.  This part takes values for $p_{\rm dark}$, $\eta_{\rm det}$
and $\eta_{\rm trans}$ for each detector. It takes the parameter
\texttt{\,vsq\,} = $|v|^2$ that is the fraction of energy tapped by the
eavesdropper.  Finally it takes a parameter for the order of R\'enyi
entropy considered in calculating the eavesdropper's
entropy. (Caution: stronger eavesdropping attacks are expected to be
implemented in the future.)  In order to speed calculations,
it calls \texttt{\,Xifn.m\,} to pre-compute values of $\Xi(\zeta,n)$, and
it also computes a list of all $n!$ for $n = 0$, \dots, \texttt{\,nmax\,}.

\vspace{5pt}\noindent (3) \texttt{\,EntPlts.m\,} plots families of curves,
such as that shown in Fig.\ 
\ref{fig:7}, working from \texttt{\,muResult\,}, obtained
either from immediately prior running of \texttt{\,muRun.m\,} or from
loading a previously saved \texttt{\,muResult\,}.

The rest of the scripts and functions are called directly or
indirectly by \texttt{\,muRun.m\,}:

\vspace{5pt}\noindent (4) \texttt{\,muScript.m\,} is a macro called three
times by \texttt{\,muRun.m\,}.

\vspace{5pt}\noindent (5) \texttt{\,Xifn.m\,} pre-computes a list of
values of $\Xi(\zeta,n)$ and stores them in \texttt{\,XiAr\,}, in order to
speed the computation of probabilities.

\vspace{5pt}\noindent (6) \texttt{\,AvEntfn.m\,} computes Eq.\
(\ref{eq:EvAvRenEFB}).

\vspace{5pt}\noindent (7) \texttt{\,Tfn.m\,} supports either
(\ref{eq:probEn2}) or (\ref{eq:probEnn2}) by computing
(\ref{eq:calTDef}) or (\ref{eq:calTEval}), respectively, as 
specified by
\texttt{\,CaseCode\,}, using
\texttt{\,Ffn.m\,}. (So far, the only energy distribution implemented is
the Poisson energy distribution, for which $C_n = e^{-\mu}\mu^n/n!$.)

\vspace{5pt}\noindent (8) \texttt{\,Ffn.m\,} computes either
(\ref{eq:CalGDef}) or (\ref{eq:calTEval}), as specified by
\texttt{\,CaseCode\,}.

\vspace{5pt}\noindent (9) \texttt{\,Gkmfn.m\,} is called by
\texttt{\,Ffn.m\,} to compute Eq.\ (\ref{eq:CalGDef}); it uses
(global)
\texttt{\,XiAr,\break CaseSetUp\,}, and
\texttt{\,gamTab\,} set up by \texttt{\,muRun.m\,}.  It calls
\texttt{\,sumfn.m\,}.

\vspace{5pt}\noindent (10) \texttt{sumfn.m} is a summing routine for efficient calculation
of sums in which the ratio of successive terms is pre-computed.

\vspace{5pt}\noindent (11) \texttt{Gnfn.m} is called by \texttt{Ffn.m} to compute Eq.\
     (\ref{eq:GnEval}).

\vspace{5pt}\noindent (12) \texttt{Gmu\_kmfn.m} computes Eq.
(\ref{eq:calGCkm}) for the special case of $C_n = e^{-\mu}\mu^n/n!$.

\vspace{5pt}\noindent (13) \texttt{Gmufn.m} computes Eq.\
(\ref{eq:calFCofL}) for the special case of $C_n = e^{-\mu}\mu^n/n!$.

\vspace{5pt}\noindent (14) \texttt{GCfn.m}. Not yet implemented.

\vspace{8pt} 
Here is the code for all these except \texttt{GCfn.m}, not yet
implemented.


\newpage\parindent=0pt
\parskip=-9pt

\centerline{\bf MATLAB PROGRAMS}

\vskip10pt
(1) \texttt{\,Partfn.m}  

\vskip6pt
\texttt{function[\,] = Partfn(nmax)}\lbrk  
\texttt{\%}  26 OCT 04\lbrk
\texttt{\%}  sets up cell for partitions from  $n  = 1$  to  $n =$ \texttt{nmax}\lbrk 
\texttt{\%}  assumes  \texttt{\,nmax} $>$  2\lbrk
\texttt{\%}  test and run for  \texttt{\,nmax = 32\,}  on 26 Oct 04 (took 20 minutes)\lbrk
\texttt{yList = cell(1,nmax);\lbrk
yList\{1\} = [1];\lbrk
yList\{2\} = [0 1;2 0];\lbrk
for jcell = 3:nmax\lbrk
yList\{jcell\} = Upfn(yList\{jcell-1\});\lbrk
end\lbrk
save Part.mat yList}\ \ \texttt{\%} ESSENTIAL RESOURCE!\lbrk

\vskip10pt
\texttt{function[y] = Upfn(Ar)}\lbrk
\texttt{\%}  Used in generating partitions of $n + 1$ from partitions of $n$;\lbrk
\texttt{\%}  \texttt{\,Ar\,} has a row for each $\nu$ vector of partitions of $n$.\lbrk
\texttt{temp = size(Ar);\lbrk
oneCol = ones(temp(1),1);\lbrk
zeroPad = zeros(size(Ar));\lbrk
oneColPadded = [oneCol zeroPad];} \texttt{\%}  row width is $n + 1$\lbrk
\texttt{zeroCol = zeros(temp(1),1);\lbrk
ArPadded = [Ar zeroCol];} \texttt{\%}  add a column to get $n + 1$ columns\lbrk
\qquad\quad     \texttt{\,ArP1 = ArPadded + oneColPadded;}\lbrk
\texttt{\%}  Up proper\lbrk
\texttt{for jRow = 1:temp(1)\lbrk
\vskip6pt
\qquad for j = 1:temp(2)\lbrk
\qquad\quad if Ar(jRow,j) > 0\lbrk
\qquad\qquad Vec = ArPadded(jRow,:);\lbrk
\qquad\qquad Vec(j) = Vec(j)-1;\lbrk
\qquad\qquad Vec(j+1) = Vec(j+1)+1;\lbrk
\qquad\qquad ArP1 = [Vec;ArP1];\lbrk
\qquad\quad end} \texttt{\%}  of \texttt{\,If}\lbrk
\qquad   \texttt{end} \texttt{\%}  of \texttt{\,For j}\lbrk
\texttt{end} \texttt{\%}  of \texttt{\,For jRow\lbrk
ArP1 = sortrows(ArP1);\lbrk
tempz = size(ArP1);\lbrk
y = ArP1(1,:);\lbrk
for jRow = 2:tempz(1)\lbrk
if ArP1(jRow-1,:)\,\,== ArP1(jRow,:)\lbrk
else\lbrk   
y = [y;ArP1(jRow,:)];\lbrk
end} \texttt{\%}  of \texttt{\,If\lbrk
end} \texttt{\%}  of \texttt{\,For jRow}  
 
\newpage
(2) \texttt{\,muRun.m}
\vskip10pt

\texttt{\%}  Compute and store and plot $\mu$-dependence of figure of merit\lbrk
\texttt{\%}  and \texttt{\,pGood}, \texttt{\,pSiftErr}, and \texttt{\,AvEnt}\lbrk
\texttt{\%}  Define figure of merit  \texttt{\,\,= pGood .*  AvEnt}\lbrk
\texttt{\%}  1 NOV 04\lbrk
\texttt{global alpha0 dark vsq CaseSetUp zetaVec XiAr NsqFac yList}\lbrk
\vskip12pt

\texttt{\%}  **********************  VARY TO DEFINE CASE  **********************\lbrk
\vskip10pt

\texttt{etaDet = [0.1 0.1 0.1 0.1] ;}\lbrk
\texttt{\%}  \texttt{etaDet = [1 1 1 1] ;\lbrk
etaTrans = [1 1 0.1 0.1];}\lbrk
\texttt{\%}  \texttt{etaTrans = [1 1 1 1];\lbrk
dark = 5.*10\^{}(-5) .* [1 1 1 1];}\lbrk
\texttt{\%}  \texttt{dark = [0 0 0 0];\lbrk
alpha0 = 1-etaDet .* etaTrans;\lbrk
vsq = .25;} \texttt{\%}  \texttt{\,vsq\,} is fraction of Bob's energy tapped by Evang.\lbrk
\texttt{\%}  \texttt{vsq = 0;\lbrk
zetaVec = [1 10 100 1000];} \texttt{\%}  Row vector \texttt{\,(1,length(zetaVec))\,} For Case\lbrk 
\texttt{Ry = 1.1;} \texttt{\%}  Order of Renyi entropy\lbrk
\texttt{\%}  can put in knob variables later\lbrk
\texttt{\%}  need to make global array of $|C_n|^2$ if cases 5, 6 used.\lbrk
\texttt{\%}  parameters for range and fineness of $\mu$\lbrk
\texttt{fine{\Large\_$\mskip1mu$}incr = .0001;\lbrk
mu{\Large\_$\mskip1mu$}begin = 0;} \texttt{\%}  may add a little to avoid problem when $p_{\rm dark}  =  0$;\lbrk 
\texttt{mu{\Large\_$\mskip1mu$}fine{\Large\_$\mskip1mu$}max = .007;\lbrk
n{\Large\_$\mskip1mu$}fine{\Large\_$\mskip1mu$}max = floor((mu{\Large\_$\mskip1mu$}fine{\Large
\_$\mskip1mu$}max - mu{\Large\_$\mskip1mu$}begin)/fine{\Large\_$\mskip1mu$}incr);\lbrk 
mu{\Large\_$\mskip1mu$}max = 0.04;\lbrk
incr = .002;} \texttt{\%}  increment $\mu$ after first steps of \texttt{\,incr/n{\Large
\_$\mskip1mu$}fine{\Large\_$\mskip1mu$}max\lbrk
n{\Large\_$\mskip1mu$}incr{\Large\_$\mskip1mu$}max = 1 + floor((mu{\Large
\_$\mskip1mu$}max - mu{\Large\_$\mskip1mu$}begin -\lbrk
n{\Large\_$\mskip1mu$}fine{\Large\_$\mskip1mu$}max*fine{\Large\_$\mskip1mu$}incr)/incr);}\lbrk
\texttt{\%}  \texttt{mu{\Large\_$\mskip1mu$}max = (n{\Large\_$\mskip1mu$}incr{\Large
\_$\mskip1mu$}max+1)*incr\lbrk
muVec = zeros(1,n{\Large\_$\mskip1mu$}incr{\Large\_$\mskip1mu$}max+n{\Large
\_$\mskip1mu$}fine{\Large\_$\mskip1mu$}max+1);} 
\texttt{\%} will be loaded with $\mu$ values\lbrk
\vskip12pt

\texttt{\%}  ************************    END OF CASE DEF    ************************\lbrk
\vskip10pt

\texttt{CaseSetUp = cell(1,6);\lbrk
CaseSetUp\{1\} = etaDet;\lbrk
CaseSetUp\{2\} = etaTrans;\lbrk
CaseSetUp\{3\} = dark;\lbrk
CaseSetUp\{4\} = vsq;\lbrk
CaseSetUp\{5\} = zetaVec;} \texttt{\%}  row vector\lbrk 
\texttt{CaseSetUp\{6\} = Ry;} \texttt{\%}  Renyi entropy of order \texttt{\,Ry\,}. load \texttt{\,Part.mat}

\vskip10pt
\texttt{nmax = 32;\lbrk
if ((zetaVec == zetaVecOld) \& (nmax == nmaxOld))\lbrk
else\lbrk
XiAr = Xifn(zetaVec,nmax,yList);} \texttt{\%}  assumes  \texttt{\,yList\,} on
hand\hfil\break
\texttt{\%}  \texttt{XiAr = cell(1,nmax+1);}\lbrk
\texttt{\%}  \texttt{cell\{n+1\}\,} is \texttt{\,Xi(zeta,n)},  a column vector \texttt{\,(1,length(zetaVec))\lbrk
zetaVecOld = zetaVec;\lbrk
nmaxOld = nmax;\lbrk
end}\lbrk
\texttt{\%}  set up table of factorials to speed calculation\lbrk
\texttt{global gamTab\lbrk
gamTab = cell(1,nmax+1);\lbrk
for kk = 1:33\lbrk
gamTab\{kk\} = factorial(kk-1);\lbrk
end}\lbrk
\texttt{\%}  END of \texttt{\,setUp.m}\lbrk

\vskip10pt
\texttt{\%}  GET LIMITS  for  \texttt{\,muMeritLims\lbrk
CaseRestore = CaseSetUp\{5\};\lbrk
CaseSetUp\{5\} = 0;\lbrk
muScript\lbrk
\qquad\quad     pGoodZ0 = pGood;\lbrk
AvEntZ0 = AvEnt;\lbrk
FigMerZ0 = FigMer;\lbrk
pSiftErrZ0 = pSiftErr;\lbrk
CaseSetUp\{5\} = 10\^{}50;\lbrk
muScript\lbrk
pGoodZInf = pGood;\lbrk
AvEntZInf = AvEnt;\lbrk
FigMerZInf = FigMer;\lbrk
pSiftErrZInf = pSiftErr;}\lbrk

\vskip10pt
\texttt{\%}  Get in-between values of  \texttt{\,zeta\lbrk 
CaseSetUp\{5\} = CaseRestore;\lbrk
muScript\lbrk
\vskip10pt
muResult = cell(1,5);\lbrk
muResult\{1\} = CaseSetUp;\lbrk
muResult\{2\} = [pGoodZ0;pGood;pGoodZInf;muVec];\lbrk
muResult\{3\} = [AvEntZ0;AvEnt;AvEntZInf;muVec];\lbrk
muResult\{4\} = [FigMerZ0;FigMer;FigMerZInf;muVec];\lbrk
muResult\{5\} = [pSiftErrZ0;pSiftErr;pSiftErrZInf;muVec];}\lbrk
\vskip10pt
\texttt{\%}  next can be run separately as \texttt{\,EntPlts\lbrk
\qquad\quad FigMerPlt = muResult\{4\};\lbrk
plot(muVec,FigMerPlt(2,:))\lbrk
\qquad\quad xlabel({\usq}mu{\usq})\lbrk
\qquad\quad ylabel({\usq}FigMerit{\usq})\lbrk
\qquad\quad title({\usq}FigMerit vs.\ mu; zeta = 0; zeta = inf --{\usq})\lbrk
\qquad\quad text(.5,.05, {\usq}$\backslash$itp{\Large\_$\mskip1mu$}\{$\backslash$rm dark\} = 0, 
$\backslash$eta{\Large\_$\mskip1mu$}\{$\backslash$rm\lbrk
det\} =  0.1,$\backslash$eta{\Large\_$\mskip1mu$}\{$\backslash$rm Trans\} = 1.{\usq})}

\newpage
\clearpage
(3) \texttt{\,EntPlts.m}

\vspace{10pt}
\texttt{\%}  Program For plotting data In  \texttt{\,muResult\,}  gotten from running  \texttt{\,muRun\,}  or\lbrk 
\texttt{\%}  loading  \texttt{\,muResult\,}\lbrk
\texttt{\%}  Plots choice of ``\texttt{\,kind\,}''.\lbrk
\texttt{\%}  6 NOVEMBER 2004\lbrk
\texttt{kind = input({\usq}1 for pGood, 2 for AvEnt, 3 for FigMer, 4 for\lbrk 
pSiftErr{\usq})}\lbrk
\texttt{\%}  takes $\mu$ values as vector  \texttt{\,muX}, an edited  \texttt{\,muVec\,}  taken from edited \texttt{\,muResult\,}\lbrk
\texttt{pltVec = muResult\{kind+1\};}\lbrk
\vspace{14pt}
\texttt{\%}  *******************   VARY UNTIL next asterisks TO EDIT\lbrk
\vspace{10pt}
\texttt{startDim = size(pltVec);}\lbrk
\vspace{8pt}
\texttt{\%} \texttt{pltVec(:,40:startDim(2)) = [\,];} \texttt{\%}  ** TEMPORARY; comment OUT\lbrk
\texttt{pltVec(:,1:20)=[\,];}\lbrk
\texttt{\%}  *********************\lbrk
\texttt{tempDim = size(pltVec);\lbrk
muX = pltVec(tempDim(1),:);} \texttt{\%}  retrieves (edited) \texttt{\,muVec\,}\lbrk
\texttt{\%}  \texttt{kind = 1\ }  gets  \texttt{\ pltVec = pGood}\lbrk
\texttt{\%}  \texttt{kind = 2\ }  gets  \texttt{\ pltVec = AvEnt}\lbrk
\texttt{\%}  \texttt{kind = 3\ }  gets  \texttt{\ pltVec = FigMer}\lbrk
\texttt{\%}  \texttt{kind = 4\ }  gets  \texttt{\ pltVec = pSiftErr\lbrk
CasePlt = muResult\{1\};} \texttt{\%}  \texttt{CaseSetUp\,}  For data In  \texttt{\,muResult\lbrk
etaDetPlt = CasePlt\{1\};\lbrk
etaTransPlt = CasePlt\{2\};\lbrk
darkPlt = CasePlt\{3\};\lbrk
vsqPlt = CasePlt\{4\};\lbrk
RyPlt = CasePlt\{6\};}\lbrk
\vspace{10pt} 
\texttt{pdarkStr =\lbrk 
[{\usq}[{\usq},num2str(darkPlt(1)),{\usq},{\usq},num2str(darkPlt(2)),{\usq},{\usq},...\lbrk
\qquad\qquad\qquad  num2str(darkPlt(3)),{\usq},{\usq},num2str(darkPlt(4)),{\usq}]{\usq}];\lbrk
etaDetStr =\lbrk 
[{\usq}[{\usq},num2str(etaDetPlt(1)),{\usq},{\usq},num2str(etaDetPlt(2)),{\usq},{\usq},...\lbrk
\qquad\qquad\qquad  num2str(etaDetPlt(3)),{\usq},{\usq},num2str(etaDetPlt(4)),{\usq}]{\usq}];\lbrkk
etaTransStr =\lbrk 
[{\usq}[{\usq},num2str(etaTransPlt(1)),{\usq},{\usq},num2str(etaTransPlt(2)),...\lbrk
\noindent {\usq},{\usq},num2str(etaTransPlt(3)),{\usq},{\usq},num2str(etaTransPlt(4)),{\usq}]{\usq}];}

\vspace{12pt}
\texttt{Titles = cell(1,4);\lbrk
\qquad\quad Titles\{1\} = {\usq}Prob.\ of correct, sifted bit vs.\ mu{\usq};}\lbrk
\texttt{\%}  factor of 1/2 For \texttt{\,prob\,}  of correct basis\lbrk
\qquad\quad    \texttt{\,Titles\{2\} = ...\lbrk 
[{\usq}Evangeline{\usq}{\usq}s  AvEnt for correct, sifted bits vs.\ mu; R =\lbrk 
\noindent {\usq},... num2str(RyPlt)];\lbrk
Titles\{3\} =  {\usq}FigMerit vs.\ mu{\usq};\lbrk
Titles\{4\} = {\usq}prob.\ of error in sifted bits vs.\ mu{\usq};\hfil\break
Ylab = cell(1,4);\lbrk
Ylab\{1\} = {\usq}pGood{\usq};\lbrk
Ylab\{2\} = {\usq}Evangeline AvEnt{\usq};\lbrk
Ylab\{3\} = {\usq}FigMerit{\usq};\lbrk
Ylab\{4\} = {\usq}pSiftErr{\usq};}\lbrk
\texttt{\%}  \texttt{plot(muX,pltVec)\lbrk
plot(muX,pltVec(1,:),muX,pltVec(2,:),{\usq}--{\usq},muX,pltVec(3,:),{\usq}-\lbrk
.{\usq},...\lbrk
muX,pltVec(4,:),{\usq}-{\usq},muX,pltVec(5,:),{\usq}-{\usq},muX,pltVec(6,:),{\usq}-{\usq})\lbrk
\qquad\quad  xlabel({\usq}mu{\usq})\lbrk
\qquad\quad  ylabel(Ylab\{kind\})\lbrk
\qquad\qquad\quad  title(Titles\{kind\})\lbrk
\qquad\quad  axis([0 max(muX) -inf inf])\lbrk 
gtext([{\usq}$\backslash$itp{\Large\_$\mskip1mu$}\{$\backslash$rm dark\} = {\usq},pdarkStr,...\lbrk
\noindent {\usq}$\backslash$it, $\backslash$eta{\Large\_$\mskip1mu$}\{$\backslash$rm det\} = {\usq},etaDetStr, 
{\usq}$\backslash$it, $\backslash$eta{\Large\_$\mskip1mu$}\{$\backslash$rm trans\} =\lbrk
\noindent {\usq}...,etaTransStr,{\usq}, vsq = {\usq},num2str(vsqPlt)])}\lbrk
\vspace{10pt}
\texttt{\%}  \texttt{muResult = cell(1,5);}\lbrk
\texttt{\%}  \texttt{muResult\{1\} = CaseSetUp;}\lbrk
\texttt{\%}  \texttt{CaseSetUp = cell(1,6);}\lbrk
\texttt{\%}  \texttt{CaseSetUp\{1\} = etaDet;}\lbrk
\texttt{\%}  \texttt{CaseSetUp\{2\} = etaTrans;}\lbrk
\texttt{\%}  \texttt{CaseSetUp\{3\} = darkPlt;}\lbrk
\texttt{\%}  \texttt{CaseSetUp\{4\} = vsq;}\lbrk
\texttt{\%}  \texttt{CaseSetUp\{5\} = zetaVec;} \texttt{\%}  row vector\lbrk 
\texttt{\%}  \texttt{CaseSetUp\{6\} = Ry;} \texttt{\%}  Renyi entropy of order \texttt{\,Ry}\lbrk
\vspace{10pt}
\texttt{\%}  \texttt{muResult\{2\} = [pGoodZ0;pGood;pGoodZInf];}\lbrk
\texttt{\%}  \texttt{muResult\{3\} = [AvEntZ0;AvEnt;AvEntZInf];}\lbrk
\texttt{\%}  \texttt{muResult\{4\} = [FigMerZ0;FigMer;FigMerZInf];}\lbrk
\texttt{\%}  \texttt{muResult\{5\} = [pSiftErrZ0;pSiftErr;pSiftErrZInf];}

\newpage
\clearpage
(4) \texttt{\,muScript.m}

\vspace{10pt}
\texttt{\%}  2 Nov 04 for use by \texttt{\,muRun.m}\lbrk
\qquad\quad \texttt{zVecLen = length(CaseSetUp\{5\});\lbrk
pGood = zeros(zVecLen,n{\Large\_$\mskip1mu$}incr{\Large\_$\mskip1mu$}max+1+n{\Large
\_$\mskip1mu$}fine{\Large\_$\mskip1mu$}max);}\lbrk 
\texttt{\%}  +1 so \texttt{\,ForLoop\,} starts at 1 with   $\mu\approx 0$\lbrk
\vspace{10pt}
\texttt{AvEnt = pGood;} \texttt{\%}  same format\lbrk
\texttt{pSiftErr = pGood;} \texttt{\%}  again same format\lbrk
\texttt{\%}  \texttt{Ry\,} order of Renyi entropy from \texttt{\,muRun}\lbrk
\texttt{\%}  Do fine steps at beginning\lbrk
\texttt{\%}  \texttt{mu = .00001}; \texttt{\%}  sloppy fix of problem with $\mu =  0$  when $p_{\rm dark} = 0$\lbrk
\texttt{mu = 0;\lbrk
\quad for kt = 1:n{\Large\_$\mskip1mu$}fine{\Large\_$\mskip1mu$}max\lbrk
\qquad t0110 = Tfn([2 mu],[0 1 1 0]);\lbrk
\qquad t1001 = Tfn([2 mu],[1 0 0 1]);\lbrk
\qquad t1010 = Tfn([2 mu],[1 0 1 0]);\lbrk
\qquad t0101 = Tfn([2 mu],[0 1 0 1]);\lbrk
\qquad pGood(:,kt) = t0110 + t1001;\lbrk
\qquad AvEnt(:,kt) = AvEntfn([2 mu],Ry);\lbrk 
\qquad pSiftErr(:,kt) = (t1010 + t0101)./(t0101+t0110+t1001+t1010);\lbrkk
\qquad FigMer = pGood .* AvEnt;\lbrk
\qquad muVec(kt) = mu;\lbrk
\qquad mu = mu+incr/n{\Large\_$\mskip1mu$}fine{\Large\_$\mskip1mu$}max;\lbrk
\quad end}\lbrk
\vspace{10pt}
\texttt{mu = mu-.0001;} \texttt{\%}  don't need fix $\mu > 0$\lbrk
\quad \texttt{for kt =
n{\Large\_$\mskip1mu$}fine{\Large\_$\mskip1mu$}max+1:n{\Large\_$\mskip1mu$}incr{\Large
\_$\mskip1mu$}max+1+n{\Large\_$\mskip1mu$}fine{\Large\_$\mskip1mu$}max\lbrk
\qquad t0110 = Tfn([2 mu],[0 1 1 0]);\lbrk
\qquad t1001 = Tfn([2 mu],[1 0 0 1]);\lbrk
\qquad t1010 = Tfn([2 mu],[1 0 1 0]);\lbrk
\qquad t0101 = Tfn([2 mu],[0 1 0 1]);\lbrk
\qquad pGood(:,kt) = t0110 + t1001;\lbrk
\qquad AvEnt(:,kt) = AvEntfn([2 mu],Ry);\lbrk 
\qquad pSiftErr(:,kt) = (t1010 + t0101)./(t0101+t0110+t1001+t1010);\lbrkk
\qquad FigMer = pGood .* AvEnt;\lbrk
\qquad muVec(kt) = mu;\lbrk
\qquad mu = mu+incr;\lbrk
\quad end}

\newpage
\clearpage
(5) \texttt{\,Xifn.m}

\vspace{10pt}
\texttt{function[XiAr] = Xifn(zeta,nmax,yList)}\lbrk
\texttt{\%}  \texttt{XiAr(m,n+1) = Xi(zeta(m),n)}\lbrk
\texttt{\%}  \texttt{yList\,} in   $\,\tilde{\ }$/matlab/qed/Part.mat\lbrk
\texttt{\%}  (Before running  \texttt{\,Xifn.\,}  load  \texttt{Part.mat})\lbrk
\texttt{\%}  1 NOV 04\lbrk
\texttt{\%}  26 October 04 Preliminary test gets OK limits\lbrk 
\texttt{\%}  30 OCT Made \texttt{\,XiAr\,}  into  \texttt{\,cell(1,nmax+1)\,}  to get enough dynamic range\lbrk
\texttt{\%}  \texttt{kappaAr(m,n) = kappa(zeta(m),n)}\lbrk
\texttt{\%}  Accepts a vector of values of zeta\lbrk
\texttt{\%}  Fails if \texttt{\,nmax > nmax\,} used in running \texttt{\,partfn.m}\lbrk
\texttt{\%}  use in loop for \texttt{\,Xi(zeta(m),0) ... Xi(zeta(m),nmax)\,} by\lbrk
\texttt{\%}  \texttt{XiAr(m,1) ... XiAr(m,nmax+1)\lbrk
zlen = length(zeta);}\lbrk
\texttt{\%}  \texttt{*XiAr = zeros(zlen,nmax+1);\lbrk
XiAr = cell(nmax+1);} \texttt{\%}  allows much bigger range of values than array does\lbrk
\texttt{XiAr\{1\} = ones(zlen,1);\lbrk
XiAr\{2\} = ones(zlen,1);\lbrk
kappaAr = kappafn(zeta,nmax);\lbrk
if nmax < 3\lbrk
return\lbrk
else\lbrk
for ncc = 3:nmax+1\lbrk
XiAr\{ncc\} = zeros(zlen,1);\lbrk
Lst = yList\{ncc-1\};\lbrk
dims = size(Lst);\lbrk
jmax = dims(1);\lbrk
for j = 1:jmax\lbrk
XiAr\{ncc\} = XiAr\{ncc\}+termfn(Lst(j,:),kappaAr);\lbrk
end} \texttt{\%}  \texttt{For j}\lbrk
\texttt{end} \texttt{\%}  \texttt{For ncc\lbrk
end} \texttt{\%}  \texttt{If nmax\lbrk
\vspace{10pt} 
function[kappaAr] = kappafn(zeta,nmax)}\lbrk
\texttt{\%}  \texttt{kappa(zeta(m),n)}\lbrk
\texttt{\%}  1 NOV 04\lbrk
\texttt{\%}  Accepts a vector of values of \texttt{\,zeta}\lbrk
\texttt{\%}  \texttt{kappaAr(m,n) = kappa(zeta(m),n)}\lbrk
\texttt{\%}  \texttt{kappa(zeta,n) = R(zeta,n)./sqrt(zeta\^{}2+1)\lbrk
x = zeta.\^{}2;\lbrk
R = zeros(nmax,length(zeta));} \texttt{\%}  will be transposed later\lbrk
\texttt{R(2,:)\,\,= ones(1,length(zeta));\lbrk
pvec = x./(4.*(x+1));\lbrk
for nct = 3:nmax\lbrk
R(nct,:)\,\,= 1./(1-pvec.*R(nct-1,:));\lbrk
end}

\newpage
\clearpage
\texttt{kappaAr = R * diag(1./sqrt(x+1));\lbrk
kappaAr(1,:)\,\,= ones(1,length(zeta));\lbrk
kappaAr = kappaAr{\usq};} \texttt{\%}  {\usq}\lbrk
\texttt{\%}  checked asymptotic \texttt{\,--> 2/(sqrt(x + 1) + 1)\,} as $n$ gets big\lbrk
\vspace{10pt}
\texttt{function[yCol] = termfn(vec,kappaAr)}\lbrk
\texttt{\%}  does \texttt{\,Column\,} over  \texttt{\,zeta(m)\,}  For one term where \texttt{\,vec\,} is a partition of $n$\lbrk
\texttt{\%}  For \texttt{\,n $\backslash$le 32}\lbrk
\texttt{\%}  using \texttt{\,kappaAr(m,n) = kappa(zeta(m),n)\lbrk
n1 = length(vec);\lbrk
dims = size(kappaAr);\lbrk
zlen = dims(1);\lbrk
yCol = ones(zlen,1);\lbrk
for j = 1:n1\lbrk
yCol = (kappaAr(:,j)./j).\^{}vec(j)/factorial(vec(j)).*yCol;\lbrk
end\lbrk
yCol = yCol.*factorial(n1);}

\newpage
\clearpage
(6) \texttt{\,AvEntfn.m}

\vspace{10pt}
\texttt{function[z{\Large\_$\mskip1mu$}out] = AvEntfn(CaseCode,R)}\lbrk
\texttt{\%}  Assumes \texttt{\,CaseCode\,} is \texttt{\,[2 mu]\,} or \texttt{\,[3 n]}\lbrk    
\texttt{\%}  29 October put in ``\texttt{if t1001 + t0110 > 0}'' to get rid of \texttt{\,0/0}; untested\lbrk
\texttt{\%}  30 OCT 2004  get \texttt{\,zeta\,} from \texttt{\,setUp}\lbrk
\texttt{global CaseSetUp\lbrk
if CaseCode(1)==2} \texttt{\%}  BIG BLOCK\lbrk
\quad \texttt{mu = CaseCode(2);\lbrk
\quad kmax = 16+3.*mu;} \texttt{\%}  from study with \texttt{\,poisfn.m}\lbrk
\texttt{\%} \texttt{numerator = 0;\lbrk
zVecLen = length(CaseSetUp\{5\});\lbrk
\qquad\quad  numerator = zeros(zVecLen,1);\lbrk
\quad for k = 0:kmax\lbrk
\qquad mmax = kmax - k;\lbrk
\qquad for m = 0:mmax\lbrk
\qquad t1001 = Tfn([5 mu k m],[1 0 0 1]);\lbrk
\qquad t0110 = Tfn([5 mu k m],[0 1 1 0]);\lbrk
\qquad if t1001+t0110 > 0\lbrk
\qquad pEv = t1001./(t1001+t0110);\lbrk 
\qquad y = log(pEv.\^{}R + (1-pEv).\^{}R)./((1-R)*log(2));\lbrk
\qquad numerator = numerator + (t1001+t0110).*y;\lbrk
\qquad end} \texttt{\%}  \texttt{if t1001+t0110...\lbrk
\qquad end} \texttt{\%}  \texttt{for m\lbrk
\quad end} \texttt{\%}  \texttt{for k\lbrk
\quad denom = Tfn([2 mu],[1 0 0 1]) + Tfn([2 mu],[0 1 1 0]);\lbrk
\quad z{\Large\_$\mskip1mu$}out = numerator./denom;\lbrk
else} \texttt{\%}  BIG BLOCK\lbrk
\quad \texttt{\%}  assume \texttt{[3 n]\lbrk
\quad n = CaseCode(2);\lbrk
\quad numerator = 0;\lbrk
\quad for k = 0:n\lbrk
\qquad for m = 0:n-k\lbrk
\qquad\quad t1001 = Tfn([6 n k m],[1 0 0 1])\lbrk
\qquad\quad t0110 = Tfn([6 n k m],[0 1 1 0])\lbrk
\qquad\quad if t1001 + t0110 > 0\lbrk
\qquad\quad pEv = t1001/(t1001+t0110);\lbrk
\qquad\quad y = log(pEv\^{}R + (1-pEv)\^{}R)/((1-R)*log(2));\lbrk
\qquad\quad numerator = numerator + (t1001+t0110).*y;\lbrk
\qquad\quad end} \texttt{\%}  \texttt{if t1001 + t0110...\lbrk
\qquad end} \texttt{\%}  \texttt{for m\lbrk
\quad end} \texttt{\%}  \texttt{for k\lbrk
\quad denom = Tfn([3 n],[1 0 0 1]) + Tfn([3 n],[0 1 1 0]);\lbrk
\quad z{\Large\_$\mskip1mu$}out = numerator./denom;\lbrk
end} \texttt{\%}  BIG BLOCK

\newpage
\clearpage
\vspace{10pt}
\texttt{\%}  \texttt{function[y{\Large\_$\mskip1mu$}out] = Tfn(CaseCode,nVec)}\lbrk
\texttt{\%}  \texttt{$\backslash$mathcal\{T\}\,} for various cases\lbrk
\texttt{\%}\texttt{\%}  \texttt{nVec\,} is negatively coded bit vector; this restricted implementation fails if \texttt{\,nVec\,}  has\lbrk 
\texttt{\%}\texttt{\%}  more than 4 zeros.\lbrk
\vspace{10pt}
\texttt{\%}  \texttt{Casecode\,} can be \texttt{\,[1 Cpt] [2 mu] [3 n] [4 Cpt k m] [5 mu k m] }\lbrk
\texttt{\%}  \texttt{[6 n k m]}\lbrk
\texttt{\%}  \texttt{Cpt\,} is a real number or integer that points to an array \texttt{\,Cvec\,} of coefficients.  
 
\newpage
\clearpage
(7) \texttt{\,Tfn.m}

\vspace{10pt}
\texttt{function[y{\Large\_$\mskip1mu$}out] = Tfn(CaseCode,nVec)}\lbrk
\texttt{\%}  \texttt{$\backslash$mathcal\{T\}\,} for various cases\lbrk
\texttt{\%}  19 OCT 04\lbrk
\texttt{\%}  \texttt{Casecode\,} can be \texttt{[1 Cpt] [2 mu] [3 n] [4 Cpt k m] [5 mu k m]}\lbrk 
\texttt{\%}  \texttt{[6 n k m]}\lbrk
\texttt{\%}  \texttt{Cpt\,} is a real number or integer that points to an array \texttt{\,Cvec\,} of coefficients.\lbrk
\vspace{10pt}
\texttt{\%}  \texttt{nVec\,} is negatively coded bit vector; this restricted implementation fails if \texttt{\,nVec\,} has more\lbrk
\texttt{\%}  than 4 zeros.\lbrk
\vspace{10pt}
\quad \texttt{ctab = [0 0 0 0; 1 0 0 0; 0 1 0 0; 1 1 0 0;...\lbrk
\hspace{5.81em}0 0 1 0; 1 0 1 0; 0 1 1 0; 1 1 1 0;...\lbrk
\hspace{5.81em}0 0 0 1; 1 0 0 1; 0 1 0 1; 1 1 0 1;...\lbrk
\hspace{5.81em}0 0 1 1; 1 0 1 1; 0 1 1 1; 1 1 1 1];}\lbrk
\vspace{8pt}
\texttt{\%}  set up \texttt{\,Index\,} as vector that shows where the zeros in \texttt{\,nVec\,} are located\lbrk
\texttt{\%}  \texttt{Ctr\,} gets incremented to the number of zeros in \texttt{\,nVec}.\lbrk
\texttt{temp = size(nVec);\lbrk
dim{\Large\_$\mskip1mu$}nVec = temp(2);\lbrk
dim{\Large\_$\mskip1mu$}Ctr = dim{\Large\_$\mskip1mu$}nVec - sum(nVec);\lbrk
jIndex = 0;\lbrk
if dim{\Large\_$\mskip1mu$}Ctr == 0} \texttt{\%}  starts big block\lbrk
\qquad\texttt{y{\Large\_$\mskip1mu$}out = (-1)\^{}sum(nVec).* Ffn(CaseCode,nVec);\lbrk
else\lbrk
\qquad y{\Large\_$\mskip1mu$}out = 0;\lbrk
\qquad Index = zeros(1,dim{\Large\_$\mskip1mu$}Ctr);\lbrk
\qquad nxtVec = nVec;\lbrk
\quad for jt = 1:dim{\Large\_$\mskip1mu$}nVec\lbrk
\qquad if nVec(jt)==0\lbrk
\qquad\quad jIndex = jIndex+1;\lbrk  
\qquad\quad Index(jIndex) = jt;\lbrk
\qquad end \texttt{\%} \texttt{If nVec}\lbrk 
\quad end} \texttt{\%} \texttt{For jt}\lbrk
\qquad\quad \texttt{\%}  \texttt{Index}  [checked and works]\lbrk
\qquad\quad \texttt{for jt = 1:2\^{}dim{\Large\_$\mskip1mu$}Ctr\lbrk
\qquad\qquad for jtt = 1:dim{\Large\_$\mskip1mu$}Ctr\lbrk
\qquad\qquad\quad nxtVec(Index(jtt)) = ctab(jt,jtt);\lbrk
\qquad\qquad end} \texttt{\%} \texttt{For jtt}\lbrk
\qquad\qquad \texttt{\%} \texttt{nxtVec} [checked and works]\lbrk
\qquad \texttt{y{\Large\_$\mskip1mu$}out = y{\Large\_$\mskip1mu$}out + Ffn(CaseCode,nxtVec);}\lbrk
\texttt{\%} \texttt{jt{\Large\_$\mskip1mu$}report = jt} \texttt{\%}  * drop in production\lbrk
\qquad\quad \texttt{end} \texttt{\%} \texttt{For jt\lbrk
end} \texttt{\%}  of big block\lbrk
\qquad \texttt{y{\Large\_$\mskip1mu$}out = (-1)\^{}sum(nVec).* y{\Large\_$\mskip1mu$}out;}

\newpage
\clearpage
(8) \texttt{\,Ffn.m}

\vspace{10pt}
\texttt{function[z{\Large\_$\mskip1mu$}out]= Ffn(CaseX,nVecX)}\lbrk
\texttt{\%}  \texttt{$\backslash$mathcal\{F\}\,} -- function for various cases.\ \ \texttt{Casecodes [1 Cpt] [2 mu] [3 n]\lbrk  
\texttt{\%}  [4 Cpt k m] [5 mu k m] [6 n k m]}\lbrk
\texttt{\%}  \texttt{Cpt\,} is a real number or integer that points to an array \texttt{\,Cvec\,} of coefficients.\lbrk  
\texttt{\%}  called by \texttt{\,Tfn.m\,} which supplies \texttt{\,CaseX\,} and \texttt{\,nVecX}.\lbrk
\texttt{global alpha0 dark vsq} \texttt{\%}  supplied by \texttt{\,muRun\,} or by \texttt{\,setUp.m\lbrk
alpha = ones(1,4);\lbrk
darkfac = 1;\lbrk
usq = 1-vsq;\lbrk
for kt = 1:4\lbrk
\qquad if nVecX(kt)==1\lbrk
\qquad\quad alpha(kt) = alpha0(kt);\lbrk
\qquad\quad darkfac = - darkfac .*(1-dark(kt));} \texttt{\%}  might vectorize later\lbrk
\qquad \texttt{end} \texttt{\%} \texttt{If nVecX\lbrk
end} \texttt{\%} \texttt{For kt\lbrk
w = alpha(1).*alpha(4).*usq;\lbrk
x = alpha(1).*vsq;\lbrk 
y = alpha(2).*alpha(3).*usq;\lbrk
z = alpha(2).*vsq;\lbrk
switch CaseX(1)\lbrk
case 1\lbrk
z{\Large\_$\mskip1mu$}out = GCfn(w,x,y,z);} \texttt{\%} NOT YET IMPLEMENTED\lbrk
\texttt{case 2\lbrk
\qquad\quad mu{\Large\_$\mskip1mu$}arg = CaseX(2);\lbrk     
z{\Large\_$\mskip1mu$}out = Gmufn(mu{\Large\_$\mskip1mu$}arg,w,x,y,z);\lbrk
case 3\lbrk
\qquad\quad n{\Large\_$\mskip1mu$}arg = CaseX(2);\lbrk
z{\Large\_$\mskip1mu$}out = Gnfn(n{\Large\_$\mskip1mu$}arg,w,x,y,z);\lbrk
case 4\lbrk 
\qquad\quad k{\Large\_$\mskip1mu$}arg = CaseX(2);\lbrk
\qquad\quad m{\Large\_$\mskip1mu$}arg = CaseX(3);\lbrk 
z{\Large\_$\mskip1mu$}out = GCkmfn(k{\Large\_$\mskip1mu$}arg,m{\Large\_$\mskip1mu$}arg,w,x,y,z);} \texttt{\%}  
assumes global \texttt{$\backslash$mathbf\{C\}\lbrk
case 5\lbrk
\qquad\quad mu{\Large\_$\mskip1mu$}arg = CaseX(2);\lbrk
\qquad\quad k{\Large\_$\mskip1mu$}arg = CaseX(3);\lbrk
\qquad\quad m{\Large\_$\mskip1mu$}arg = CaseX(4);\lbrk
z{\Large\_$\mskip1mu$}out = Gmu{\Large\_$\mskip1mu$}kmfn(mu{\Large\_$\mskip1mu$}arg,k{\Large 
\_$\mskip1mu$}arg,m{\Large\_$\mskip1mu$}arg,w,x,y,z);\lbrk
case 6\lbrk
\qquad\quad n{\Large\_$\mskip1mu$}arg = CaseX(2);\lbrk
\qquad\quad k{\Large\_$\mskip1mu$}arg = CaseX(3);\lbrk
\qquad\quad m{\Large\_$\mskip1mu$}arg = CaseX(4);\lbrk
\qquad\quad z{\Large\_$\mskip1mu$}out = Gnkmfn(n{\Large\_$\mskip1mu$}arg,k{\Large
\_$\mskip1mu$}arg,m{\Large\_$\mskip1mu$}arg,w,x,y,z);\lbrk
otherwise\lbrk
end\lbrk
z{\Large\_$\mskip1mu$}out = z{\Large\_$\mskip1mu$}out .* darkfac;}

\newpage
\clearpage
(9) \texttt{\,Gkmfn.m}

\vspace{12pt}
\texttt{function[y{\Large\_$\mskip1mu$}out] = Gnkmfn(n,k,m,w,x,y,z)}\lbrk
\texttt{\%}  4 NOV 04\lbrk
\texttt{\%}  ``vectorize'' over \texttt{\,w, x, y, z\,} as well as over \texttt{\,zetaX}\lbrk
\texttt{\%}  works only if \texttt{\,32 $\backslash$ge n $\backslash$ge k + m}\lbrk
\texttt{\%}  implemented to assume \texttt{\,zeta = $\backslash$infty if zeta $\backslash$ge 10\^{}40}\lbrk
\texttt{\%}  \texttt{gamTab\{n+1\} = factorial(n)} For \texttt{\,0 $\backslash$le n $\backslash$le 32} (think gamma function).\lbrk
\vspace{10pt}
\texttt{global XiAr CaseSetUp gamTab\lbrk 
zetaX = CaseSetUp\{5\};\lbrk
if zetaX==0} \texttt{\%}  BIG BLOCK\lbrk
\qquad\qquad \texttt{a0 = (gamTab\{n-k+1\}/(gamTab\{n-k-m+1\}*gamTab\{m+1\}*(n+1)))...\lbrkk
.*x.\^{}k.*z.\^{}m.*y.\^{}(n-k-m);\lbrk
\qquad if n-k-m==0\lbrk
\qquad\qquad y{\Large\_$\mskip1mu$}out = a0;\lbrk
\qquad else\lbrk
\qquad\quad avec = ones(1,n-k-m+1);\lbrk
\qquad\quad avec(1) = a0;\lbrk
\qquad\qquad for jt =1:n-k-m\lbrk
\qquad\qquad\qquad avec(jt+1) = (jt+k)*(n-k-m+1-jt)/((n-k+1-jt)*jt);\lbrk
\qquad\qquad end} \texttt{\%}  \texttt{For jt\lbrk
\qquad\quad xt = w./y;\lbrk
\qquad\quad y{\Large\_$\mskip1mu$}out = sumfn(avec,xt);\lbrk
\qquad end} \texttt{\%}  \texttt{If n-k-m==0\lbrk
else if zetaX < 10\^{}40} \texttt{\%}  General Case\lbrk
\qquad \texttt{zlen = length(zetaX);\lbrk
\qquad NsqFac = zeros(zlen,1);\lbrk
\quad for jt = 0:n\lbrk
\qquad NsqFac = NsqFac + XiAr\{jt+1\}.*XiAr\{n-\lbrk
jt+1\}./(gamTab\{jt+1\}*gamTab\{n-jt+1\});\lbrk
\quad end} \texttt{\%}  \texttt{For jt\lbrk
\qquad NsqFac = (gamTab\{n+1\}).*NsqFac;\lbrk
\qquad NsqFac = 1./NsqFac;\lbrk
\qquad Const = NsqFac.*(gamTab\{n+1\}/(gamTab\{m+1\}*gamTab\{k+1\}))*...\lbrk
\qquad (x.\^{}k.*z.\^{}m.*y.\^{}(n-k-m));\lbrk
\qquad\qquad\quad Tot = zeros(zlen,length(w));\lbrk
\qquad\qquad\quad for j = 0:n-k-m\lbrk
\qquad\qquad\quad Tot = Tot + XiAr\{j+k+1\}.*XiAr\{n-k-j+1\}*(w./y).\^{}j./...\lbrk
\qquad\qquad\quad (gamTab\{j+1\}*gamTab\{n-k-m-j+1\});\lbrk
\qquad\qquad\quad end} \texttt{\%}  \texttt{For j\lbrk
\qquad\quad y{\Large\_$\mskip1mu$}out = Tot.*Const;}\lbrk
\qquad \texttt{else} \texttt{\%}  \texttt{limit as zeta --> $\backslash$infty\lbrk
\qquad\qquad\qquad y{\Large\_$\mskip1mu$}out = 2\^{}(-n)*gamTab\{n+1\}.*x.\^{}k.*z.\^{}m .*(w+y).\^{}(n-k-\lbrk
\qquad\qquad\qquad m)./...\lbrk
\qquad\qquad\qquad (gamTab\{k+1\}*gamTab\{m+1\}*gamTab\{n-k-m+1\});\lbrk
\qquad\quad end\lbrk
end} \texttt{\%}  BIG BLOCK

\newpage
\clearpage
(10) \texttt{\,sumfn.m} 

\vspace{10pt}
\texttt{function[y] = sumfn(ar,x)}\lbrk
\qquad\quad \texttt{\%}  \texttt{ar\,} is a vector of the form
\texttt{\,[a{\Large\_$\mskip1mu$}0,a{\Large\_$\mskip1mu$}1/a{\Large\_$\mskip1mu$}0,a{\Large
\_$\mskip1mu$}2/a{\Large\_$\mskip1mu$}1,...,a{\Large\_$\mskip1mu$}n/a{\Large\_$\mskip1mu$}\{n-1\}]}\lbrk
\texttt{\%}  \texttt{x\,} is a variable value or a vector of variable values\lbrk
\texttt{\%}  \texttt{y = sum{\Large\_$\mskip1mu$}\{j=0\}\^{}n a{\Large\_$\mskip1mu$}j.*x.\^{}j.\lbrk
dimar = size(ar);}\lbrk
\vspace{10pt}
\texttt{ntemp = dimar(2);} \texttt{\%}  biggest $n$ is \texttt{\,ntemp - 1};\lbrk
\texttt{\%}  set up \texttt{\,ztemp\lbrk
dimx = size(x);\lbrk
\qquad\quad nxtemp = dimx(2);\lbrk
ztemp = ones(1,nxtemp);\lbrk
for ct = ntemp:-1:2} \texttt{\%}  trouble if \texttt{\,ntemp < 2}.\lbrk
\texttt{y = x.*ar(ct).*ztemp;\lbrk
ztemp = 1+y;\lbrk
end\lbrk
y = ar(1).*ztemp;}\lbrk
\texttt{\%}  test by \texttt{\,ar = [2 1.2 1.2 1.2 1.2], x = [2 3 4]}\lbrk
\texttt{\%}  \texttt{sumfn([2 1.2 1.2 1.2 1.2],[2 3 4])\,} Checks.

\newpage
\clearpage
(11) \texttt{\,Gnfn.m}

\vspace{10pt}
\texttt{function[y{\Large\_$\mskip1mu$}Gn] = Gnfn(n{\Large\_$\mskip1mu$}arg,w,x,y,z)}\lbrk
\texttt{\%}  26 OCT 04 \texttt{\%}  does limit of \texttt{\,zeta --> $\backslash$infty\,} if \texttt{\,zetaX > 10\^{}40}\lbrk
\texttt{\%}  29 OCT tested general \texttt{\,zetaX\,} against limits OK.\lbrk
\texttt{\%}  Old \texttt{\,Gnfn(zetaX,n{\Large\_$\mskip1mu$}arg,w,x,y,z)\,}\lbrk
\texttt{global XiAr CaseSetUp gamTab\lbrk
x1 = (y+z);\lbrk
x2 = (w+x);\lbrk
zetaX = CaseSetUp\{5\};\lbrk
if zetaX == 0\lbrk
\quad if x1 == x2\lbrk
\qquad y{\Large\_$\mskip1mu$}Gn = x1.\^{}n{\Large\_$\mskip1mu$}arg;\lbrk
\quad else\lbrk 
\qquad y{\Large\_$\mskip1mu$}Gn = (x1.\^{}(n{\Large\_$\mskip1mu$}arg+1)-x2.\^{}(n{\Large
\_$\mskip1mu$}arg+1))./((n{\Large\_$\mskip1mu$}arg+1).*(x1-x2));\lbrk
\quad end} \texttt{\%} \texttt{If x1\lbrk
else\lbrk 
\quad if zetaX < 10\^{}40\lbrk
\qquad\quad zlen = length(zetaX);\lbrk
\qquad\quad NsqFac = zeros(zlen,1);\lbrk
\qquad for jt = 0:n{\Large\_$\mskip1mu$}arg\lbrk
\qquad\quad NsqFac = NsqFac + XiAr\{jt+1\}.*XiAr\{n{\Large\_$\mskip1mu$}arg-jt+1\}./...\lbrk
\qquad\quad (gamTab\{jt+1\}*gamTab\{n{\Large\_$\mskip1mu$}arg-jt+1\});\lbrk
\qquad end} \texttt{\%}  \texttt{For jt\lbrk
\qquad\quad NsqFac = (gamTab\{n{\Large\_$\mskip1mu$}arg+1\}).*NsqFac;\lbrk
\qquad\quad NsqFac = 1./NsqFac;\lbrk
\qquad\qquad Const = NsqFac*(y+z).\^{}n{\Large\_$\mskip1mu$}arg;}\lbrk
\vspace{10pt}   
\texttt{\%}  Modified from \texttt{\,Gnkmfn\lbrk
\qquad\qquad Tot = zeros(zlen,length(w));\lbrk
\qquad\quad for j = 0:n{\Large\_$\mskip1mu$}arg\lbrk
\qquad\qquad Tot = Tot + XiAr\{j+1\}.*XiAr\{n{\Large\_$\mskip1mu$}arg-j+1\}*...\lbrk
\qquad\qquad ((w+x)./(y+z)).\^{}j./(gamTab\{j+1\}*gamTab\{n{\Large\_$\mskip1mu$}arg-j+1\});}\lbrk 
\texttt{\%}   (*)\lbrk
\qquad\quad \texttt{end} \texttt{\%}  \texttt{For j\lbrk
\qquad\qquad Tot = gamTab\{n{\Large\_$\mskip1mu$}arg+1\}.*Tot;\lbrk
\qquad y{\Large\_$\mskip1mu$}Gn = Tot.*Const;\lbrk
\quad else} \texttt{\%}  limit as \texttt{\,zeta --> $\backslash$infty\lbrk
\qquad y{\Large\_$\mskip1mu$}Gn = ((x1+x2)./2).\^{}n{\Large\_$\mskip1mu$}arg;\lbrk
\quad end} \texttt{\%}  \texttt{\,If zetaX < 10\^{}40\lbrk
end}

\newpage
\clearpage
(12) \texttt{\,Gmu{\Large\_$\mskip1mu$}kmfn.m}
 
\vspace{10pt}
\texttt{function[y{\Large\_$\mskip1mu$}ans] =
Gmu{\Large\_$\mskip1mu$}kmfn(mu{\Large\_$\mskip1mu$}arg,k{\Large\_$\mskip1mu$}arg,m{\Large
\_$\mskip1mu$}arg,w,x,y,z)}\lbrk
\texttt{\%}  30 OCT 04\lbrk
\texttt{global CaseSetUp gamTab\lbrk
zetaX = CaseSetUp\{5\};}\lbrk
\texttt{\%}  checked sum against \texttt{\,Gmufn}, OK.\lbrk
\quad \texttt{if zetaX < 10\^{}40} \texttt{\%}  \texttt{zeta $\backslash$ge 10\^{}40\,} treated as infinite\lbrk
\qquad \texttt{nmax = floor(3*mu{\Large\_$\mskip1mu$}arg+16);\lbrk
\qquad y{\Large\_$\mskip1mu$}ans = 0;\lbrk
\qquad\quad for nt = k{\Large\_$\mskip1mu$}arg + m{\Large\_$\mskip1mu$}arg:nmax\lbrk
\qquad incr = Gnkmfn(nt,k{\Large\_$\mskip1mu$}arg,m{\Large\_$\mskip1mu$}arg,w,x,y,z);\lbrk
\qquad\qquad\quad y{\Large\_$\mskip1mu$}ans = y{\Large\_$\mskip1mu$}ans + mu{\Large
\_$\mskip1mu$}arg\^{}nt.*incr./gamTab\{nt+1\};\lbrk
\qquad\quad end} \texttt{\%} \texttt{For nt\lbrk
\qquad y{\Large\_$\mskip1mu$}ans = exp(-mu{\Large\_$\mskip1mu$}arg).*y{\Large\_$\mskip1mu$}ans;\lbrk
\quad else} \texttt{\%}  limit as \texttt{\,zeta --> $\backslash$infty\lbrk 
\qquad mu2 = mu{\Large\_$\mskip1mu$}arg/2;\lbrk
\qquad y{\Large\_$\mskip1mu$}ans = exp(-mu2.*(2-w-y)).*(mu2.*x).\^{}k{\Large
\_$\mskip1mu$}arg.*(mu2.*z).\^{}m{\Large\_$\mskip1mu$}arg./...\lbrkk
\qquad (gamTab\{k{\Large\_$\mskip1mu$}arg+1\}*gamTab\{m{\Large\_$\mskip1mu$}arg+1\});\lbrk
end}

\newpage
\clearpage
(13) \texttt{\,Gmufn.m} 

\vspace{10pt}
\texttt{function[y{\Large\_$\mskip1mu$}out] = Gmufn(mu{\Large\_$\mskip1mu$}arg,w,x,y,z);\lbrk
global CaseSetUp gamTab}\lbrk
\texttt{\%}  \texttt{gamTab\{n+1\} = factorial\{n\}\,} For \texttt{\,n = 0, \dots, nmax\lbrk
zetaX = CaseSetUp\{5\};\lbrk
if zetaX == 0\lbrk
\qquad x1 = (y+z);\lbrk
\qquad x2 = (w+x);\lbrk
\quad if mu{\Large\_$\mskip1mu$}arg == 0\lbrk
\qquad y{\Large\_$\mskip1mu$}out = 1;\lbrk
\quad else\lbrk
\quad if x1 == x2\lbrk
\qquad y{\Large\_$\mskip1mu$}out = exp(-mu{\Large\_$\mskip1mu$}arg.*(1-x1));\lbrk
\quad else\lbrk
\qquad y{\Large\_$\mskip1mu$}out = exp(-mu{\Large\_$\mskip1mu$}arg).*(exp(mu{\Large
\_$\mskip1mu$}arg.*x1)\lbrk
\qquad\qquad\qquad\quad -exp(mu{\Large\_$\mskip1mu$}arg.*x2))./...(mu{\Large
\_$\mskip1mu$}arg.*(x1-x2));\lbrk
\quad end} \texttt{\%}  \texttt{\,If x1 == x2\lbrk
\quad end} \texttt{\%}  \texttt{\,If mu{\Large\_$\mskip1mu$}arg == 0\lbrk
else if zetaX > 10\^{}40\lbrk
\qquad y{\Large\_$\mskip1mu$}out = exp(-mu{\Large\_$\mskip1mu$}arg.*(2-w-x-y-z)./2);\lbrk
else} \texttt{\%}  General Case\ \ \texttt{0 < zetaX $\backslash$le 10\^{}40}\lbrk
\qquad \texttt{\,y{\Large\_$\mskip1mu$}out = zeros(length(zetaX),length(w));}\lbrk 
\texttt{\%}  column vector of same length as \texttt{\,\,zetaX\lbrk
\qquad nmax = floor(3*mu{\Large\_$\mskip1mu$}arg+16);\lbrk
\quad for nt = 0:nmax\lbrk
\qquad incr = Gnfn(nt,w,x,y,z);\lbrk
\qquad y{\Large\_$\mskip1mu$}out = y{\Large\_$\mskip1mu$}out + mu{\Large
\_$\mskip1mu$}arg\^{}nt.*incr./gamTab\{nt+1\};}\lbrk 
\texttt{\%}  \texttt{\,gamTab\{nt+1\} = factorial(nt)\lbrk
\quad end} \texttt{\%} \texttt{For nt\lbrk
\qquad y{\Large\_$\mskip1mu$}out = exp(-mu{\Large\_$\mskip1mu$}arg).*y{\Large
\_$\mskip1mu$}out;\lbrk
end} \texttt{\%}  General Case\lbrk
\texttt{end}

\end{document}